\documentclass[english,aps,prd,nofootinbib,eqsecnum,superscriptaddress]{revtex4-2}
\usepackage[T1]{fontenc}
\usepackage[utf8]{luainputenc}
\usepackage{amsmath}
\usepackage{amssymb}
\usepackage{graphicx}
\usepackage{setspace}

\makeatletter

\newcommand{\lyxdot}{.}

\usepackage{esint}
\usepackage[marginal]{footmisc}
\usepackage{braket}

\usepackage{verbatim}
\usepackage{fancyhdr}
\allowdisplaybreaks[4]

\DeclareMathSizes{12}{12}{7}{7}
\renewcommand{\theequation}{\arabic{section}.\arabic{equation}}

\def\frontmatter@abstractheading{\centerline{\textbf{Abstract}}}
\g@addto@macro\abstract{\ignorespaces}

\makeatother

\usepackage{babel}
\begin{document}
\title{Phonon-mediated  Migdal effect in semiconductor detectors}
\author{Zheng-Liang Liang}
\email{liangzl@mail.buct.edu.cn}

\affiliation{College of Mathematics and Physics, Beijing University of Chemical
Technology~\\
Beijing 100029, China}
\author{Chongjie Mo}
\email{cjmo@csrc.ac.cn}

\affiliation{Beijing Computational Science Research Center, Beijing, 100193, China}
\author{Fawei Zheng}
\email{fwzheng@bit.edu.cn}

\affiliation{Centre for Quantum Physics, Key Laboratory of Advanced~\\
 Optoelectronic Quantum Architecture and Measurement(MOE),~\\
School of Physics, Beijing Institute of Technology, Beijing, 100081,
China}
\author{Ping Zhang}
\email{pzhang2012@qq.com}

\affiliation{School of Physics and Physical Engineering, Qufu Normal University~\\
 Qufu, 273165, China}
\affiliation{Institute of Applied Physics and Computational Mathematics~\\
Beijing, 100088, China}
\begin{abstract}
\setstretch{1.2}The Migdal effect inside detectors provides a new possibility of probing
the sub-GeV dark matter (DM) particles. While there has been well-established
methods treating the Migdal effect in isolated atoms, a coherent and
complete description of the valence electrons in semiconductor is
still absent. The bremstrahlung-like approach is a promising attempt,
but it turns invalid for DM masses below a few tens of MeV. In this
paper, we lay out a framework where phonon is chosen as an effective
degree of freedom to describe the Migdal effect in semiconductors.
In this picture, a valence electron is excited to the conduction state
via exchange of a virtual phonon, accompanied by a multi-phonon process
triggered by an incident DM particle. Under the incoherent approximation,
it turns out that this approach can effectively push the sensitivities
of the semiconductor targets further down to the MeV DM mass region.
\end{abstract}
\maketitle

\section{Introduction}

The search for low-mass dark matter~(DM) particle has progressed
tremendously in the past decade, with significant theoretical and
experimental advances in new detection channels and materials~\citep{Kahn:2021ttr},
such as in semiconductors~\citep{Essig:2011nj,Graham:2012su,Essig:2015cda,Hochberg:2016sqx},
Dirac materials~\citep{Hochberg:2017wce,Coskuner:2019odd,Geilhufe:2019ndy},
superconductors~\citep{Hochberg:2015pha,Hochberg:2016ajh}, superfluid
helium~\citep{Knapen2017,Caputo:2019cyg,Caputo:2019xum}, and via
phonon excitations~\citep{Griffin:2018bjn,Knapen2018,Campbell-Deem:2019hdx}
and bremsstrahlung photons~\citep{Kouvaris:2016afs,Bell:2019egg},
as well as other proposals and analyses~\citep{Essig:2012yx,Lee:2015qva,Hochberg:2015fth,Bloch:2016sjj,Derenzo:2016fse,Hochberg:2016ntt,Essig:2016crl,Kadribasic:2017obi,Essig:2017kqs,Arvanitaki:2017nhi,Budnik:2017sbu,SHARMA2017326,Cavoto:2017otc,Liang:2018bdb,Heikinheimo:2019lwg,Trickle:2019nya,Trickle:2019ovy,Catena:2019gfa,Andersson:2020uwc,Trickle:2020oki,Griffin:2020lgd,Chen:2022pyd,Hamaide:2021hlp,Chao:2021liw}.

The Migdal effect has attracted wide interest recently because the
study in Ref.~\citep{Ibe2018} has shown that in theory the suddenly
struck nucleus can produce ionized electrons more easily than anticipated
for an incident sub-GeV DM particle, so exploring relevant parameter
region is plausible for the present detection technologies. Although
the Migdal effect has not been directly observed in a nuclear collision,
attempts to make the first such measurement from neutron-nucleus scattering
are underway~\citep{Nakamura:2020kex,MIGDALcollab,OHare:2022jnx}.
After Ref.~\citep{Ibe2018}, there has emerged numerous theoretical
proposals~\citep{Ibe2018,Dolan:2017xbu,Essig:2019xkx,PhysRevD.102.043007,Bell:2019egg,Knapen:2020aky,Liang:2020ryg,GrillidiCortona:2020owp,Liu:2020pat,Flambaum:2020xxo,Dey:2020sai,Wang:2021oha}
and experimental efforts dedicated to detecting the sub-GeV DM particles
via the Migdal effect in liquids~\citep{Aprile:2019jmx}, and in
condensed matter targets~\citep{Liu:2019kzq,Arnaud:2020svb,COSINE-100:2021poy,SuperCDMS:2022kgp}. 

Compared with the typical ionization energy thresholds in atoms $\varepsilon_{g}\sim\mathcal{O}\left(10\right)\,\mathrm{eV}$,
semiconductor targets have a much lower thresholds $\varepsilon_{g}\sim\mathcal{O}\left(1\right)\,\mathrm{eV}$,
which makes them ideal materials for further exploiting the the Migdal
effect in the probe of light DM particles. However, generalizing the
boosting argument in isolated atoms proposed in Ref.~\citep{Ibe2018}
to the crystalline environments faces both conceptual and technical
obstacles: while one keeps pace with the recoiling nucleus, the ion
lattice background will move in opposite direction, which brings no
substantial convenience in mitigating the original complexity. Thus
the semiconductor target at rest is still a preferred frame of reference.
In Ref.~\citep{PhysRevD.102.043007}, we made a tentative effort
to describe the Migdal effect in semiconductors using the tight-binding
approximation, where a Galilean boost operator is imposed specifically
onto the recoiled ion to account for the highly local impulsive effect
caused by the collision with an incident DM particle, while the extensive
nature of the electrons in solids is reflected in the hopping integrals.
Refs.~\citep{Knapen:2020aky,Liang:2020ryg} managed to describe the
Migdal effect in solids in a manner analogous to bremsstrahlung calculation,
where the valence electron is excited to the conduction state via
the bremsstrahlung photons emitted by the recoiling ion. 

The bremsstrahlung-like approach is an effective description of the
Migdal event rates for DM masses $m_{\chi}\geq50\,\mathrm{MeV}$~\citep{Knapen:2020aky}.
However, below this mass, the picture of a recoiling ion in the solid
begins to break down and the effects of phonons become important.
In Refs.~\citep{Liang:2020ryg} we proposed that the Migdal effect
in solids can alternatively be described by treating the phonon as
the mediator for the Coulomb interaction in the lattice between the
abruptly recoiling ion and itinerant electrons. Thus the objective
of this work is to provide a complete and self-contained theoretical
foundation for this idea. Within this framework, numerous phonons,
rather than an on-shell ion, are produced from the DM-nucleus scattering,
especially in the low energy regime, where the scattering is coherent
over the whole crystal. In the large momentum transfer limit however,
the recoiling on-shell ion is expected to reappear as a wave packet
supported by a large number of phonons. Such an asymptotic behavior
should self-consistently justify the impulse approximation adopted
in the bremsstrahlung-like approach. While the multi-phonon process
has been thoroughly discussed in literatures~(e.g., Ref.~\citep{Schober2014}
and references therein, and see Refs.~\citep{Kahn:2020fef,Knapen:2020aky,Berghaus:2021wrp,Campbell-Deem:2022fqm}
for recent discussions related to Migdal effect and DM searches),
the fresh idea in this paper is to incorporate the generation of phonons,
and the excitation of the electron-hole pairs, as well as the medium
effect in solids, into a common framework. By doing so, it is no longer
necessary to match the bremsstrahlung-like calculation onto the phonon
regime, and the inherent conflict between the picture of a recoiling
ion and that of the scattered phonons can be resolved altogether. 

For convenience, our discussions are carried out by using the machinery
of the \textit{quantum field theory}~(QFT), a language more familiar
to the particle physics community. This approach proves intuitive
and effective. As an interesting example, we derive the Debye-Waller
factor with the Feynman diagram method, circumventing the awkward
techniques associated with the operator commutator algebra~(see Appendix~\ref{subsec:DM-phonon}).
Based on the calculated Migdal excitation event rates using this \textit{phonon-mediated}
description, we are able to push the sensitivities of the semiconductor
detectors down to the MeV DM mass range.

This paper is organized as follows. We begin Sec.~\ref{sec:Multiphonon}
by giving the QFT framework for the multi-phonon process induced by
DM particles. Based on this discussion, we then generalize the formalism
to the Migdal excitation process in Sec.~\ref{sec:MigdalEffect}.
We conclude and make some comments on the methodology in Sec.~\ref{sec:Conclusions}.
A short review on the electrons and phonons in the context of the
QFT, as well as other supporting materials are provided in Appendix~\ref{sec:Phonons_appendix}.

\section{\label{sec:Multiphonon}multi-phonon process}

In this section we first derive the formula for the scattering cross
section between a DM particle and the target material, and then discuss
the asymptotic behavior of the phonon spectrum towards the large momentum
transfer limit. For simplicity, here we only consider the case of
the monatomic simple crystal at $0$~K.

\begin{figure}[h]
\begin{centering}
\includegraphics[scale=0.5]{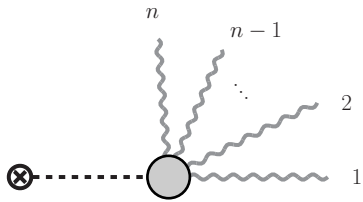}
\par\end{centering}
\caption{\label{fig:multiphonon}The diagram of the process $\chi\left(p_{\chi}\right)+\mathrm{target}\rightarrow\chi\left(p_{\chi}'\right)+\mathrm{target}+\left(\mathbf{k}_{1},\alpha_{1}\right)+\left(\mathbf{k}_{2},\alpha_{2}\right)+\cdots\left(\mathbf{k}_{n},\alpha_{n}\right)$.
See text for details.}
\end{figure}

We consider the scattering process where $n$ phonons~$\left\{ \mathbf{k}_{j},\alpha_{j}\right\} ,\,\left(j=1,\,2,\cdots,\,n\right)$
are generated by an incident DM particle in the context of the QFT,
where $\left\{ \mathbf{k}_{j}\right\} $ and $\left\{ \alpha_{j}\right\} $
represent the phonon wavevectors in the\textit{ first Brillouin zone}~(1BZ),
and phonon polarization branches, of the final states, respectively.
The relevant diagram is shown in Fig.~\ref{fig:multiphonon}, where
the initial~($\mathbf{p}{}_{\chi}$) and final~($\mathbf{p}'{}_{\chi}$)
DM states are replaced with an external source. With such replacement
it is convenient to switch the scattering theory at zero-temperature
to the linear response theory at a finite temperature, where the interest
is focused on the response of target material to external perturbations.
A more complete treatment of the composite lattice at a finite temperature
lies beyond the scope of this work, and will be pursued in further
investigation. Using the Feynman rules summarized in Appendix.~\ref{subsec:Feynman-rules},
the amplitude is read as
\begin{eqnarray}
i\mathcal{M} & = & \left(-i\right)V_{\chi N}\left(\mathbf{q}\right)N\,\sum_{\mathbf{G}}\delta_{\sum_{j}\mathbf{k}_{j}+\mathbf{q},\,\mathbf{G}}\,e^{-W\left(\mathbf{q}\right)}\prod_{j=1}^{n}\left(\frac{-i\mathbf{q}\cdot\mathbf{\boldsymbol{\epsilon}}_{\mathbf{k}_{j},\alpha_{j}}}{\sqrt{2\,N\,m_{N}\omega_{\mathbf{k}_{j},\alpha_{j}}}}\right),
\end{eqnarray}
where $\mathbf{q}=\mathbf{p}'_{\chi}-\mathbf{p}{}_{\chi}$ is the
momentum transferred to the DM particle, $\mathbf{G}$'s are reciprocal
lattice vectors, $N$ is the number of the unit cells in the crystal,
which equals the number of the atoms in a monatomic simple crystal,
$V$ is the volume of the material, $m_{N}$ is the nucleus mass,
$\boldsymbol{\epsilon}_{\mathbf{k}_{j},\alpha_{j}}$ and $\omega_{\mathbf{k}_{j},\alpha_{j}}$
are the phonon eigenvector and the eigenfrequency of branch $\alpha_{j}$
at wavevector $\mathbf{k}_{j}$, respectively; $V_{\chi N}\left(\mathbf{q}\right)$
represents the DM-nucleus contact interaction $V_{\chi N}\left(\mathbf{x}\right)$
in momentum space, which connects to the DM-nucleon cross section
$\sigma_{\chi n}$ through
\begin{eqnarray}
\left|V_{\chi N}\left(\mathbf{q}\right)\right|^{2} & = & \left|\int\mathrm{d}^{3}x\,e^{-i\mathrm{\mathbf{q}\cdot\mathbf{x}}}\,V_{\chi N}\left(\mathbf{x}\right)\right|^{2}\nonumber \\
 & = & \frac{A^{2}\pi\sigma_{\chi n}}{\mu_{\chi n}^{2}},\label{eq:Asquared}
\end{eqnarray}
with $A$ being the atomic number of the target nucleus, and $\mu_{\chi n}=m_{n}\,m_{\chi}/\left(m_{n}+m_{\chi}\right)$
representing the reduced mass of the DM~($\chi$)-nucleon~($n$)
pair system. $W\left(\mathbf{q}\right)=\sum_{\mathbf{k},\alpha}\frac{\left|\mathbf{q}\cdot\mathbf{\boldsymbol{\epsilon}}_{\mathbf{k},\alpha}\right|^{2}}{4Nm_{N}\omega_{\mathbf{k},\alpha}}$
is the Debye-Waller factor at zero-temperature. Since the lattice
is not perfectly rigid, the Debye-Waller factor accounts for the effect
of the quantum and thermal uncertainties of the positions of the nuclei
in the scattering. At $T=0\,\mathrm{K}$, only the zero-point fluctuation
is relevant. Thus, the total cross section of the DM-target scattering
is expressed as
\begin{eqnarray}
\sigma & = & \frac{2\pi}{V}N^{2}\sum_{\mathbf{q}}\frac{\left|V_{\chi N}\left(\mathbf{q}\right)\right|^{2}}{v}\sum_{\left\{ \mathbf{k}_{j},\alpha_{j}\right\} }\frac{1}{n!}\prod_{j=1}^{n}\left(\frac{\left|\mathbf{q}\cdot\mathbf{\boldsymbol{\epsilon}}_{\mathbf{k}_{j},\alpha_{j}}\right|^{2}}{2\,N\,m_{N}\omega_{\mathbf{k}_{j},\alpha_{j}}}\right)e^{-2W\left(\mathbf{q}\right)}\,\sum_{\mathbf{G}}\delta_{\sum_{j}\mathbf{k}_{j}+\mathbf{q},\,\mathbf{G}}\,\delta\left(\sum_{j=1}^{n}\omega_{\mathbf{k}_{j},\alpha_{j}}+\omega_{p'p}\right),\nonumber \\
\label{eq:crossSection0}
\end{eqnarray}
where $\omega_{p'p}=\left|\mathbf{p}'_{\chi}\right|^{2}/\left(2m_{\chi}\right)-\left|\mathbf{p}{}_{\chi}\right|^{2}/\left(2m_{\chi}\right)$
is the energy transferred to the DM particle, and $v$ is its incident
velocity. Note that the sum $\sum_{\left\{ \mathbf{k}_{j},\alpha_{j}\right\} }$
runs over all possible phonon vibration modes as the final states.
In the above expression, the integration of the out-going DM momentum
$\mathbf{p}'_{\chi}$ is traded for that over the transferred momentum
$\mathbf{q}$. Since there are $n$ identical phonons in a final state,
the integration over momenta is divided by $n!$. A convenient correspondence
$\sum_{\mathbf{G},\mathbf{G}'}\delta_{\sum_{j}\mathbf{k}_{j}+\mathbf{q},\,\mathbf{G}}\delta_{\sum_{j}\mathbf{k}_{j}+\mathbf{q},\,\mathbf{G}'}$$\sim\sum_{\mathbf{G},\mathbf{G}'}\delta_{\sum_{j}\mathbf{k}_{j}+\mathbf{q},\,\mathbf{G}}\delta_{\mathbf{G}\,\mathbf{G}'}$$\sim\sum_{\mathbf{G}}\delta_{\sum_{j}\mathbf{k}_{j}+\mathbf{q},\,\mathbf{G}}$
is adopted in evaluating the amplitude squared. A detailed discussion
on the quantization of vibrations in solids using the path integral
approach is arranged in Appendix.~\ref{sec:Phonons_appendix}.

\begin{figure}
\begin{centering}
\includegraphics[scale=0.6]{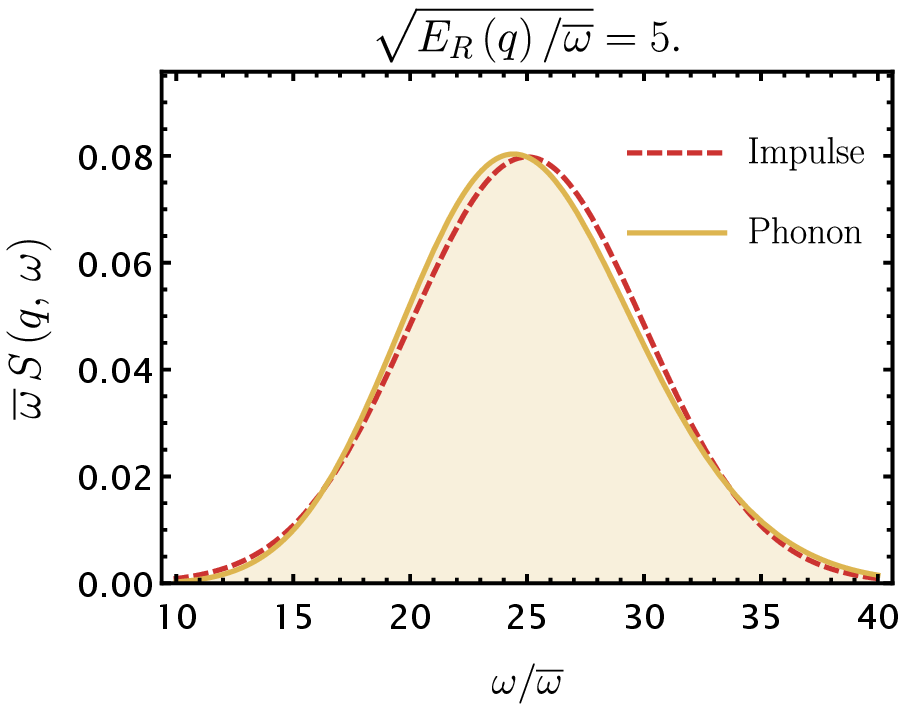}\hspace*{0.5cm}\includegraphics[scale=0.6]{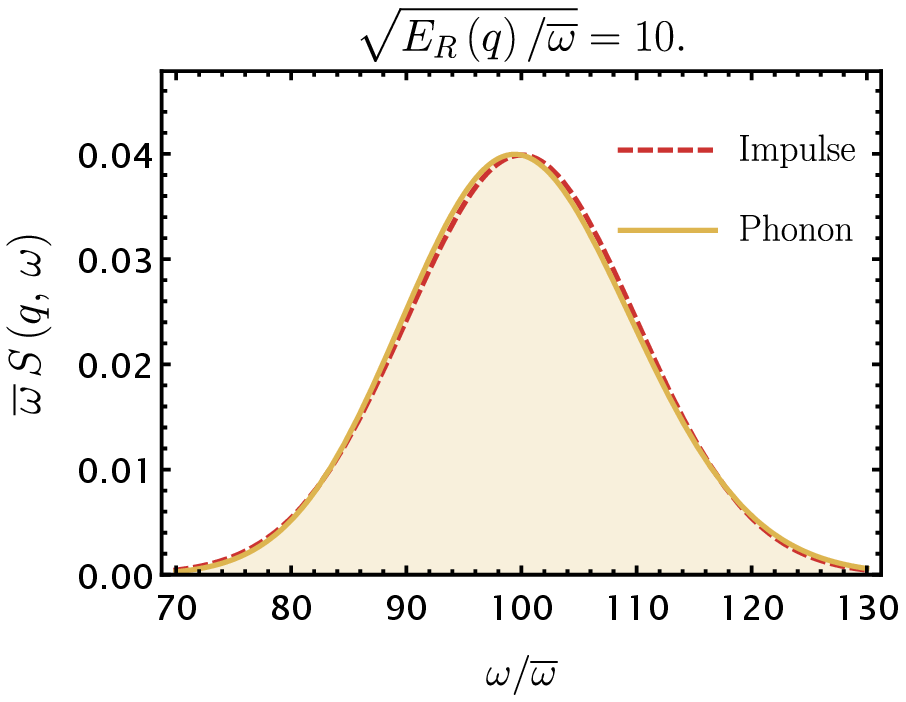}\vspace{0.7cm}
\par\end{centering}
\begin{centering}
\includegraphics[scale=0.58]{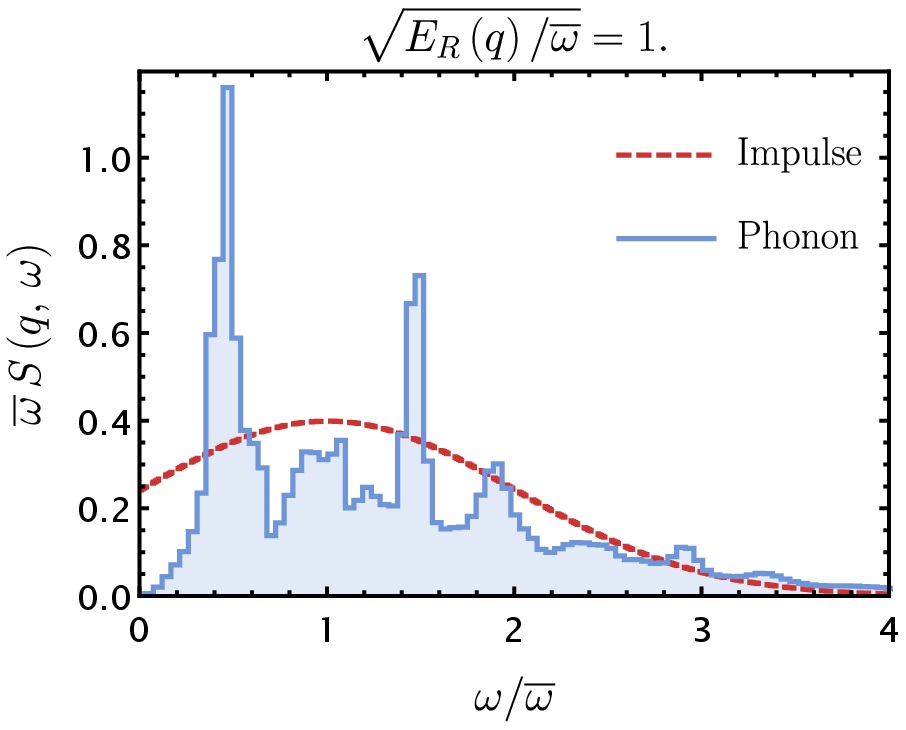}\hspace*{0.5cm}\includegraphics[scale=0.6]{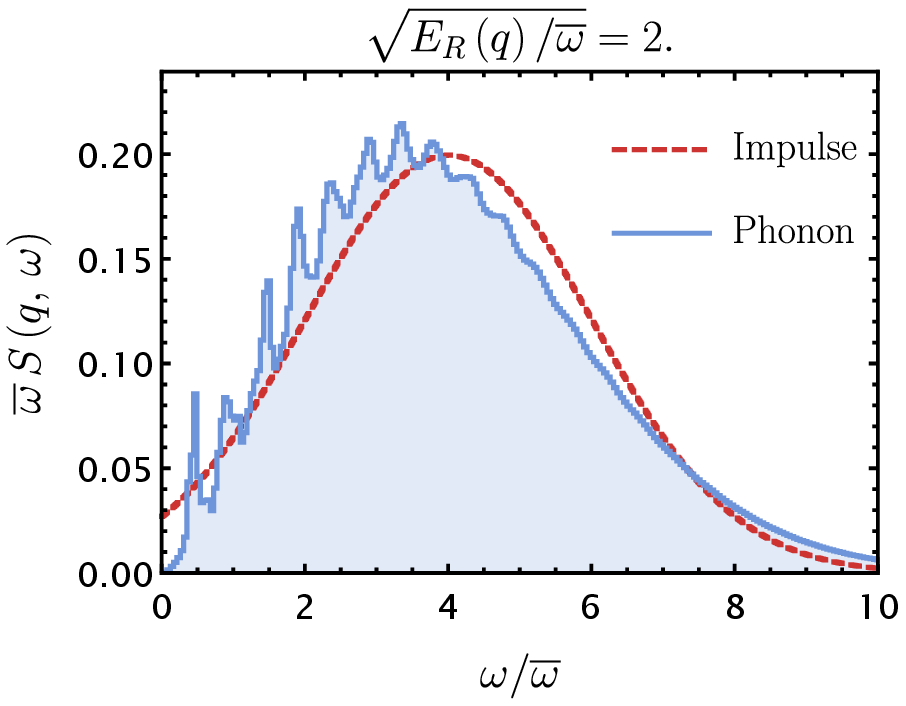}\hspace*{0.5cm}\includegraphics[scale=0.6]{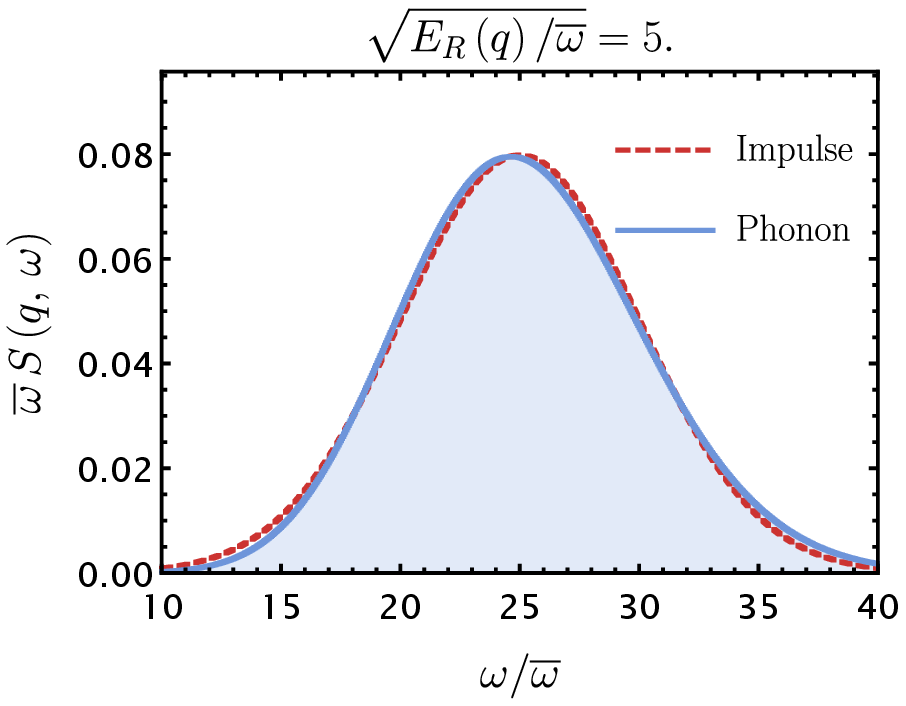}
\par\end{centering}
\caption{\label{fig:Gaussian} \textbf{\textit{Top}}: Comparisons between the
function $\overline{\omega}S\left(q,\,\omega\right)$ of silicon for
the multi-phonon distribution~(\textit{orange}) and the impulse Gaussian~(\textit{red
dashed} curves) for the ratios $\sqrt{E_{R}\left(q\right)/\overline{\omega}}=5$
and $10$, respectively. The multi-phonon distributions are estimated
with the asymptotic expansion~(up to $\mathcal{O}\left(\sqrt{\overline{\omega}/E_{R}\left(q\right)}\right)$)
presented in Eq.~(\ref{eq:Asymptotic}). \textbf{\textsl{Bottom}}:
Comparisons between the function $\overline{\omega}S\left(q,\,\omega\right)$
of silicon for the multi-phonon distribution~(\textit{blue} histograms)
obtained by the numerical recursive method and the impulse Gaussian~(\textit{red
dashed }curves) for the ratios $\sqrt{E_{R}\left(q\right)/\overline{\omega}}=1,\,2,$
and $5$, respectively. Similar discussion can be found in Ref.~\citep{Knapen:2020aky}.
See text for details.}
\end{figure}

Moreover, note that the momentum $\mathbf{q}$ can be uniquely separated
into certain reciprocal lattice $\mathbf{G}_{\mathbf{q}}$, and a
remainder part $\left[\mathbf{q}\right]$ within the 1BZ, such that
$\mathbf{q}=\mathbf{G}_{\mathbf{q}}+\left[\mathbf{q}\right]$, and
thus the summation over $\mathbf{q}$ can be equivalently expressed
as the sum $\sum_{\mathbf{G}_{\mathbf{q}}}\sum_{\left[\mathbf{q}\right]\in1\mathrm{BZ}}$.
The integration over $\left[\mathbf{q}\right]$ can always be integrated
out from the sum $\sum_{\mathbf{G}}\delta_{\sum_{j}\mathbf{k}_{j}+\mathbf{q},\,\mathbf{G}}$
for an arbitrary set of $\left\{ \mathbf{k}_{j}\right\} $ without
noticeably affecting the values of other integrand functions $\left(\cdots\right)_{\mathbf{q}}$
that are coarsely dependent on $\mathbf{q}$. The variation of the
integrand over the 1BZ is expected to be irrelevant as long as the
momentum transfer $q=\left|\mathbf{q}\right|$ is much larger than
the length of the 1BZ, i.e., $q\gg\mathcal{O}\left(1\right)\,\mathrm{keV}$.
In this case, one has the following \textit{incoherent} scattering
approximation,
\begin{eqnarray}
\sum_{\mathbf{q}}\left(\cdots\right)_{\mathbf{q}}\sum_{\mathbf{G}}\delta_{\sum_{i}\mathbf{k}_{i}+\mathbf{q},\,\mathbf{G}} & = & \sum_{\mathbf{G}_{\mathbf{q}}}\sum_{\left[\mathbf{q}\right]\in1\mathrm{BZ}}\left(\cdots\right)_{\mathbf{G}_{\mathbf{q}}+\left[\mathbf{q}\right]}\sum_{\mathbf{G}}\delta_{\sum_{i}\mathbf{k}_{i}+\left[\mathbf{q}\right],\,\mathbf{G}}\nonumber \\
 & = & \sum_{\mathbf{G}_{\mathbf{q}}}\left(\cdots\right)_{\mathbf{G}_{\mathbf{q}}+\mathbf{k}_{0}}\nonumber \\
 & \simeq & \frac{1}{N}\sum_{\mathbf{q}}\left(\cdots\right)_{\mathbf{q}},
\end{eqnarray}
where a unique $\mathbf{k}_{0}\in1\mathrm{BZ}$ satisfies $\sum_{\mathbf{G}}\delta_{\sum_{i}\mathbf{k}_{i}+\mathbf{k}_{0},\,\mathbf{G}}=1$.
This approximation amounts to smoothing out $\mathbf{q}$ within the
1BZ as if one can only see a momentum transfer with a resolution comparable
to the length of a reciprocal lattice. Next, we further approximate
that the simple lattice is isotropic. In this case, eigenenergy $\omega_{\mathbf{k},\alpha}$
remains invariant under any rotational operation $\mathcal{O}_{R}$
acting on wavevector $\mathbf{k}$, while $\mathbf{\boldsymbol{\epsilon}}_{\mathbf{k},\alpha}$
also transforms as a vector under the same $\mathcal{O}_{R}$, and
thus one has
\begin{eqnarray}
\sum_{\mathbf{k},\alpha}\frac{\left|\mathbf{q}\cdot\mathbf{\boldsymbol{\epsilon}}_{\mathbf{k},\alpha}\right|^{2}}{2Nm_{N}\omega_{\mathbf{k},\alpha}} & = & E_{R}\left(q\right)\sum_{i=1}^{3N}\frac{1}{3N}\frac{1}{\omega_{i}},\label{eq:contraction}
\end{eqnarray}
where $E_{R}\left(q\right)=q^{2}/\left(2m_{N}\right)$. This result
also holds for a monatomic cubic system~\citep{Schober2014}. In
the right-hand-side of Eq.~(\ref{eq:contraction}), we relabel the
eigenmodes $\left\{ \mathbf{k},\alpha\right\} $ with a single notation
$\left\{ i\right\} $ for brevity, and Eq.~(\ref{eq:crossSection0})
in the incoherent approximation is recast as
\begin{eqnarray}
\sigma & \simeq & \frac{2\pi}{V}N\sum_{\mathbf{q}}\frac{\left|V_{\chi N}\left(\mathbf{q}\right)\right|^{2}}{v}\times S\left(q,\,-\omega_{p'p}\right)\nonumber \\
 & = & \frac{2\pi}{V}\sum_{\mathbf{q}}N\frac{\left|V_{\chi N}\left(\mathbf{q}\right)\right|^{2}}{v}\nonumber \\
 &  & \times\sum_{\left\{ n_{i}\right\} }\frac{e^{-\frac{E_{R}\left(q\right)}{3N\omega_{1}}}}{n_{1}!}\left(\frac{E_{R}\left(q\right)}{3N\omega_{1}}\right)^{n_{1}}\cdots\frac{e^{-\frac{E_{R}\left(q\right)}{3N\omega_{3N}}}}{n_{3N}!}\left(\frac{E_{R}\left(q\right)}{3N\omega_{3N}}\right)^{n_{3N}}\delta\left(\sum_{i=1}^{3N}n_{i}\omega_{i}+\omega_{p'p}\right)\nonumber \\
 & = & \frac{2\pi}{V}\sum_{\mathbf{q}}N\frac{\left|V_{\chi N}\left(\mathbf{q}\right)\right|^{2}}{v}\times e^{-E_{R}\left(q\right)\sum_{i=1}^{3N}\frac{1}{3N}\frac{1}{\omega_{i}}}\sum_{n=0}^{+\infty}\frac{E_{R}\left(q\right)^{n}}{n!}T_{n}\left(-\omega_{p'p}\right),\label{eq:Compound}
\end{eqnarray}
where the scattering function $S\left(q,\,-\omega_{p'p}\right)$ is
defined in the third line, $n_{i}$ represents the occupation number
of the energy $\omega_{i}$, and $\overline{\omega}=\sum_{i=1}^{3N}\frac{\omega_{i}}{3N}$
is the phonon frequency averaged over the density of states~(DoS).
Note that $\sum_{\left\{ n_{i}\right\} }\frac{e^{-\frac{E_{R}\left(q\right)}{3N\omega_{1}}}}{n_{1}!}\left(\frac{E_{R}\left(q\right)}{3N\omega_{1}}\right)^{n_{1}}\cdots\frac{e^{-\frac{E_{R}\left(q\right)}{3N\omega_{3N}}}}{n_{3N}!}\left(\frac{E_{R}\left(q\right)}{3N\omega_{3N}}\right)^{n_{3N}}$
is a combined Poisson distribution, so the key problem is to determine
the probability density of the random variable $\omega=\sum_{i=1}^{3N}n_{i}\omega_{i}$
for this distribution. While it is difficult to derive an analytical
expression on a general basis, one can prove that the factor $S\left(q,\,\omega\right)$
converges to a Gaussian form in the large $q$ region, \textit{i.e}.,
$e^{-\frac{\left(\omega-E_{R}\left(q\right)\right)^{2}}{2E_{R}\left(q\right)\overline{\omega}}}/\sqrt{2\pi E_{R}\left(q\right)\overline{\omega}}$,
by using an argument analogous to that used in the proof of the central
limit theorem. Additionally, this Gaussian form converges to the delta
function in terms of the energy conservation towards the large $q$
limit, which validates the impulse approximation. In the large $q$
regime, the scattering function $S\left(q,\,\omega\right)$ can be
approximated with an asymptotic expansion with respect to parameter
$\sqrt{\overline{\omega}/E_{R}\left(q\right)}$ as follows~(see Appendix~\ref{subsec:Asymptotic}
for a detailed discussion),
\begin{eqnarray}
S\left(q,\,\omega\right) & = & \frac{e^{-\frac{\left(\omega-E_{R}\left(q\right)\right)^{2}}{2E_{R}\left(q\right)\overline{\omega}}}}{\sqrt{2\pi E_{R}\left(q\right)\overline{\omega}}}\left\{ 1+\frac{1}{6}\left(\frac{\overline{\omega^{2}}}{\overline{\omega}^{2}}\right)\cdot\left[\frac{\left(\omega-E_{R}\left(q\right)\right)^{2}}{E_{R}\left(q\right)\overline{\omega}}-3\right]\cdot\frac{\left(\omega-E_{R}\left(q\right)\right)}{\sqrt{E_{R}\left(q\right)\overline{\omega}}}\cdot\sqrt{\frac{\overline{\omega}}{E_{R}\left(q\right)}}\right\} +o\left(\sqrt{\frac{\overline{\omega}}{E_{R}\left(q\right)}}\right),\nonumber \\
\label{eq:Asymptotic}
\end{eqnarray}
with $\overline{\omega^{2}}=\sum_{i=1}^{3N}\frac{1}{3N}\omega_{i}^{2}$.
To get some sense, taking silicon target for example, in the top row
of Fig.~\ref{fig:Gaussian} we show the non-dimensional function
$\overline{\omega}S\left(q,\,\omega\right)$ for parameters $\sqrt{E_{R}\left(q\right)/\overline{\omega}}=5$
and $10$, respectively. It is evident that in the regime $\sqrt{E_{R}\left(q\right)/\overline{\omega}}\gtrsim5$,
the compound Poisson distribution in the third line of Eq.~(\ref{eq:Compound})
already well resembles the asymptotic Gaussian form in shape, except
for a minor displacement of the central value $E_{R}\left(q\right)$.

For a small transferred momentum $q$, the asymptotic expansion above
is no longer valid. In this case, one can alternatively utilize the
functions $\left\{ T_{n}\left(\omega\right)\right\} $ in the last
line of Eq.~(\ref{eq:Compound}) so as to calculate the scattering
function $S\left(q,\,\omega\right)$ in a numerical fashion. Defined
as
\begin{eqnarray}
T_{n}\left(\omega\right) & = & \frac{1}{2\pi}\int_{-\infty}^{+\infty}f\left(t\right)^{n}e^{-i\omega t}\mathrm{d}t,
\end{eqnarray}
where 
\begin{eqnarray}
f\left(t\right) & = & \sum_{i=1}^{3N}\frac{1}{3N}\frac{e^{i\omega_{i}t}}{\omega_{i}},
\end{eqnarray}
these $\left\{ T_{n}\left(\omega\right)\right\} $ can be determined
by following an iterative procedure~(see Appendix~\ref{subsec:Iterative_Multiphonon}
for further details):
\begin{eqnarray}
\int_{-\infty}^{+\infty}T_{1}\left(\omega-\omega'\right)T_{n-1}\left(\omega'\right)\mathrm{d}\omega' & = & T_{n}\left(\omega\right).
\end{eqnarray}
 In the bottom row of Fig.~\ref{fig:Gaussian} we present the non-dimensional
function $\overline{\omega}S\left(q,\,\omega\right)$ of silicon target
computed with recursive method for parameters $\sqrt{E_{R}\left(q\right)/\overline{\omega}}=1,\,2,$
and $5$, respectively. It illustrates the transition from the multi-phonon
spectrum into a Gaussion form with an increasing momentum transfer
$q$. Taking typical semiconductors such as silicon for instance,
where $\overline{\omega}=40.3\,\mathrm{meV}$, the condition $\sqrt{E_{R}\left(q\right)/\overline{\omega}}\gtrsim1$
translates to a momentum transfer $q\gtrsim\mathcal{O}\left(10\right)\,\mathrm{keV},$
which still guarantees the validity of the incoherent approximation.
In the limit $q\rightarrow\infty$, the width of the Gaussian becomes
much smaller than the central value $E_{R}\left(q\right)$, and hence
the Gaussian reduces to the $\delta$-function $\delta$$\left(\omega_{p'p}+E_{R}\left(q\right)\right)$.
Then inserting Eq.~(\ref{eq:Asquared}), and taking the correspondence
$\sum_{\mathbf{q}}\sim\frac{V}{\left(2\pi\right)^{3}}\int\mathrm{d^{3}}q$,
the cross section in the limit $q\rightarrow\infty$ becomes 
\begin{eqnarray}
\sigma & = & \frac{A^{2}\pi\sigma_{\chi n}N}{\mu_{\chi n}^{2}v}\int\frac{\mathrm{d^{3}}q}{\left(2\pi\right)^{2}}\,\delta\left(\frac{q^{2}}{2\mu_{\chi N}}+\mathbf{q}\cdot\mathbf{v}\right),
\end{eqnarray}
 which, as expected, is equal to the sum of $N$ incoherent DM-nucleus
cross sections for monatomic simple crystal structure.

\section{\label{sec:MigdalEffect}Migdal effect as a multi-phonon process}

\begin{figure}
\begin{centering}
\includegraphics[scale=0.5]{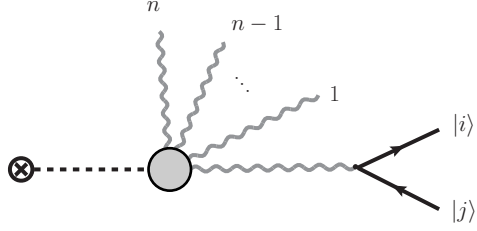}
\par\end{centering}
\caption{\label{fig:Migdal}The diagram for the Migdal effect, where an electron-hole
pair is generated, \textit{i.e.}, an electron is elevated from a valence
state $\ket{j}$ to a conduction state $\ket{i}$, via the exchange
of a virtual phonon, along with multiple on-shell phonons generated
by the DM external field.}
\end{figure}

The prospect of describing the Migdal effect in terms of phonons and
electrons has been originally sketched out in Ref.~\citep{Liang:2020ryg}.
Here we explore this approach in more details. The Migdal excitation
process is illustrated in Fig.~\ref{fig:Migdal}, where an electron-hole
pair is excited through a virtual phonon, along with a bunch of on-shell
ones produced from the collision with a DM particle. In essence, the
electron-phonon interaction reflects the Coulomb forces between the
distorted ion lattice and the itinerant electrons, for which we provide
a short review in Appendix.~\ref{subsec:Phonon-electron}. 

Following the Feynman rules summarized in Sec.~\ref{subsec:Feynman-rules},
one can read off the amplitude for the process illustrated in Fig.~\ref{fig:Migdal},
\begin{eqnarray}
i\mathcal{M} & = & \left(-i\right)V_{\chi N}\left(\mathbf{q}\right)N\,e^{-W\left(\mathbf{q}\right)}\,\prod_{s=1}^{n}\left(\frac{-i\mathbf{q}\cdot\mathbf{\boldsymbol{\epsilon}}_{\mathbf{k}_{s},\alpha_{s}}}{\sqrt{2\,N\,m_{N}\omega_{\mathbf{k}_{s},\alpha_{s}}}}\right)\nonumber \\
 &  & \times\sum_{\mathbf{G}'}\sum_{\mathbf{k},\alpha}\left(\frac{-i\mathbf{q}\cdot\mathbf{\boldsymbol{\epsilon}}_{\mathbf{k},\alpha}}{\sqrt{Nm_{N}}}\right)\left[\frac{i}{\left(\varepsilon_{i}-\varepsilon_{j}\right)^{2}-\omega_{\mathbf{k},\alpha}^{2}}\right]\left[\frac{i\,\left(\mathbf{k}+\mathbf{G}'\right)\cdot\mathbf{\boldsymbol{\epsilon}}_{\mathbf{k},\alpha}}{\sqrt{Nm_{N}}}\right]\left[\frac{-iNZ_{\mathrm{ion}}4\pi\alpha_{e}}{V\left|\mathbf{k}+\mathbf{G}'\right|^{2}}\right]\braket{i|e^{i\left(\mathbf{k}+\mathbf{G}'\right)\cdot\hat{\mathbf{x}}}|j}\nonumber \\
 &  & \times\sum_{\mathbf{G}}\delta_{\sum_{s}\mathbf{k}_{s}+\mathbf{q}+\mathbf{k},\,\mathbf{G}},\label{eq:Amplitude_Electron_phonon}
\end{eqnarray}
where $Z_{\mathrm{ion}}$ is the number of the valence electrons of
the material atom, $\alpha_{e}$ is the fine structure constant, and
$\varepsilon_{i}$~($\varepsilon_{j}$) denotes the energy of the
conduction~(valence) state $\ket{i}$~($\ket{j}$). The $n$-phonon
sector in the first line has been thoroughly discussed in the preceding
section. In the second line lies a phonon mediator with its two ends
linking the multi-phonon blob and the bare phonon-electron vertex.
For typical semiconductors, the band gaps $\varepsilon_{g}\sim\mathcal{O}\left(1\right)\mathrm{eV}$
are much larger than the phonon eigenenergies $\omega_{\mathbf{k},\alpha}\sim\mathcal{O}\left(10^{-2}\right)\mathrm{eV}$,
so the term in second line can be reduced to 
\begin{eqnarray}
 &  & \frac{Z_{\mathrm{ion}}}{\left(\varepsilon_{i}-\varepsilon_{j}\right)^{2}}\,\sum_{\mathbf{G}'}\sum_{\mathbf{k}}\left[\frac{\mathbf{q}\cdot\left(\mathbf{k}+\mathbf{G}'\right)}{m_{N}}\right]\left[\frac{4\pi\alpha_{e}}{V\left|\mathbf{k}+\mathbf{G}'\right|^{2}}\right]\braket{i|e^{i\left(\mathbf{k}+\mathbf{G}'\right)\cdot\hat{\mathbf{x}}}|j}.
\end{eqnarray}
In the derivation, we use the contraction relation $\sum_{\alpha}\left(\mathbf{q}\cdot\mathbf{\boldsymbol{\epsilon}}_{\mathbf{k},\alpha}\right)\left[\left(\mathbf{k}+\mathbf{G}'\right)\cdot\mathbf{\boldsymbol{\epsilon}}_{\mathbf{k},\alpha}\right]=\mathbf{q}\cdot\left(\mathbf{k}+\mathbf{G}'\right)$.
Recall that in Refs.~\citep{Knapen:2020aky,Liang:2020ryg} the same
amplitude is obtained in the the soft limit, \textit{i.e.}, $\mathbf{p}_{N}\cdot\left(\mathbf{k}+\mathbf{G}\right)/m_{N}\ll\varepsilon_{i}-\varepsilon_{j}$
and $\left|\mathbf{k}+\mathbf{G}\right|\ll\left|\mathbf{p}_{N}\right|$,
with $\mathbf{p}_{N}$ being the momentum of the recoiling nucleus.
In the low-energy regime however, neither the soft approximation nor
the concept of a freely-recoiling nucleus still holds. In contrast,
in the context of electrons and phonons, this particular form of amplitude
naturally extends to the low-energy regime. The excitation rate~(in
the incoherent approximation) can then be written as
\begin{eqnarray}
R & = & \frac{\rho_{\chi}}{m_{\chi}}\int\mathrm{d^{3}}v\,\sigma\,vf_{\chi}\left(\mathbf{v}\right)\nonumber \\
 & = & \frac{\rho_{\chi}}{m_{\chi}}\frac{2\pi A^{2}\sigma_{\chi n}\,Z_{\mathrm{ion}}^{2}\,N_{T}}{V^{2}\mu_{\chi n}^{2}}\int\frac{\mathrm{d}\omega}{\omega^{4}}\,\int\mathrm{d^{3}}v\,f_{\chi}\left(\mathbf{v}\right)\sum_{\mathbf{q}}\,e^{-E_{R}\left(q\right)\sum_{k}\frac{1}{3N}\frac{1}{\omega_{k}}}\,\sum_{n=0}^{+\infty}\frac{E_{R}\left(q\right)^{n}}{n!}T_{n}\left(-\mathbf{v}\cdot\mathbf{q}-\frac{q^{2}}{2m_{\chi}}-\omega\right)\nonumber \\
 &  & \times\sum_{\mathbf{G},\mathbf{G}'}\sum_{\mathbf{k}\in1\mathrm{BZ}}\left[\frac{\mathbf{q}\cdot\left(\mathbf{k}+\mathbf{G}\right)}{m_{N}}\right]\left[\frac{\mathbf{q}\cdot\left(\mathbf{k}+\mathbf{G}'\right)}{m_{N}}\right]\frac{4\pi\alpha_{e}}{\left|\mathbf{\mathbf{k}}+\mathbf{G}\right|\left|\mathbf{\mathbf{k}}+\mathbf{G}'\right|}\nonumber \\
 &  & \times\frac{2}{V}\frac{4\pi^{2}\alpha_{e}}{\left|\mathbf{\mathbf{k}}+\mathbf{G}\right|\left|\mathbf{\mathbf{k}}+\mathbf{G}'\right|}\sum_{i,j}\braket{i|e^{i\left(\mathbf{\mathbf{\mathbf{\mathbf{k}}+\mathbf{G}}}'\right)\cdot\hat{\mathbf{x}}}|j}\braket{j|e^{-i\left(\mathbf{\mathbf{\mathbf{\mathbf{k}}+\mathbf{G}}}\right)\cdot\hat{\mathbf{x}}}|i}\delta\left(\varepsilon_{i}-\varepsilon_{j}-\omega\right),\label{eq:eventRate}
\end{eqnarray}
where $\rho_{\chi}$ represents the DM local density, $f_{\chi}\left(\mathbf{v}\right)$
is the DM velocity distribution. Note that the number of the nuclei
$N$ in solids, which equals the number of the primitive cells for
a simple lattice structure, is explicitly represented with $N_{T}$
here. The factor $2$ in the last line counts the two spin orientations
for each valence state. For the present we have not taken into account
the renormalization effect in our discussion, which can displace the
locations of the phonon poles, and induce the screening of the Coulomb
interaction. Since the band gaps are far larger than the phonon eigenenergies,
only the screening effect that leads to a reduction of the scattering
rate is relevant for our discussion. 

Here we take the homogeneous electron gas~(HEG) for a schematic illustration.
As shown in Appendix.~\ref{subsec:Renormalization} and explained
in Ref.~\citep{Liang:2020ryg,Knapen:2020aky,Knapen:2021run}, the
screening of the electron-phonon vertex adds an inverse dielectric
function $\epsilon^{-1}\left(\mathbf{k},\,\omega\right)$ to the amplitude
analogous to Eq.~(\ref{eq:Amplitude_Electron_phonon}) of a crystal
structure, while the last line in Eq.~(\ref{eq:eventRate}) corresponds
to $\mathrm{Im}\left[\epsilon\left(\mathbf{k},\,\omega\right)\right]$
at the \textit{random phase approximation}~(RPA) level. Therefore,
the overall screening effect is encoded in the \textit{energy loss
function}~(ELF) $\mathrm{Im}\left[-\epsilon^{-1}\left(\mathbf{k},\,\omega\right)\right]=\mathrm{Im}\left[\epsilon\left(\mathbf{k},\,\omega\right)\right]/\left|\epsilon\left(\mathbf{k},\,\omega\right)\right|^{2}$,
which right approaches the last line in Eq.~(\ref{eq:eventRate})
in the limit $\alpha_{e}\rightarrow0$, as the screening effect becomes
negligible. Using the substitution $\sum_{\mathbf{q}}\sim\frac{V}{\left(2\pi\right)^{3}}\int\mathrm{d^{3}}q$
and $\sum_{\mathbf{k}}\sim\frac{V}{\left(2\pi\right)^{3}}\int_{1\mathrm{BZ}}\mathrm{d^{3}}k$,
the above event rate can be recast as
\begin{eqnarray}
R & = & \frac{\rho_{\chi}}{m_{\chi}}\frac{2\,\alpha_{e}A^{2}\sigma_{\chi n}\,Z_{\mathrm{ion}}^{2}\,N_{T}}{3\,\Omega\,\mu_{\chi n}^{2}\,m_{N}^{2}}\,\int\mathrm{d^{3}}v\,\frac{f_{\chi}\left(\mathbf{v}\right)}{v}\int\frac{\mathrm{d}\omega}{\omega^{4}}\,\mathcal{F}\left(\omega\right)\,\int q^{3}\,\mathrm{d}q\,e^{-E_{R}\left(q\right)\sum_{i}\frac{1}{3N}\frac{1}{\omega_{i}}}\,\sum_{n=0}^{+\infty}\frac{E_{R}\left(q\right)^{n}}{n!}\int_{0}^{qv-\frac{q^{2}}{2m_{\chi}}-\omega}T_{n}\left(E\right)\mathrm{d}E,\nonumber \\
\label{eq:eventRate-1}
\end{eqnarray}
where the nondimensional factor 
\begin{widetext}
\begin{eqnarray}
\mathcal{F}\left(\omega\right) & = & \sum_{\mathbf{G},\mathbf{G}'}\int_{1\mathrm{BZ}}\frac{\varOmega\,\mathrm{d}^{3}k}{\left(2\pi\right)^{3}}\,\frac{\left(\mathbf{k}+\mathbf{G}\right)\cdot\left(\mathbf{k}+\mathbf{G}'\right)}{\left|\mathbf{k}+\mathbf{G}\right|\left|\mathbf{k}+\mathbf{G}'\right|}\,\mathrm{Im}\left[-\widetilde{\epsilon}_{\mathbf{G},\mathbf{G}'}^{-1}\left(\mathbf{k},\omega\right)\right]\label{eq:crystal form factor}
\end{eqnarray}
represents the averaged energy loss function, with $\Omega$ being
the volume of the unit cell, and $\mathrm{Im}\left[\tilde{\epsilon}_{\mathbf{G},\mathbf{G}'}^{-1}\left(\mathbf{\mathbf{\mathbf{k}}},\omega\right)\right]$~(see
Appendix~\ref{subsec:RPA}) being the EFL for the crystal structure.
$\mathcal{F}\left(\omega\right)$ has been calculated for diamond
and silicon targets in Ref.~\citep{Liang:2020ryg}. Eq.~(\ref{eq:eventRate-1})
applies for the crystal targets that can be considered as isotropic,
in which case only the one-dimensional DM speed distribution is relevant
for the calculation of the excitation rate in Eq.~(\ref{eq:eventRate}),
and thus an isotropic velocity distribution $f_{\chi}\left(\mathbf{v}\right)$
is assumed. In the derivation, we first integrate out the angular
variable of velocity $\mathbf{v}$ with respect to $\mathbf{q}$,
which converts to the integral over variable $E$ in Eq.~(\ref{eq:eventRate-1}),
and restore the full integration over velocity distribution~(by adding
a factor $1/2$) for convenience. Then we integrate out the solid
angle of momentum transfer $\mathbf{q}$ using the Legendre addition
theorem, which leads to the factor $\mathcal{F}\left(\omega\right)$.
\begin{figure}
\begin{centering}
\includegraphics[scale=0.72]{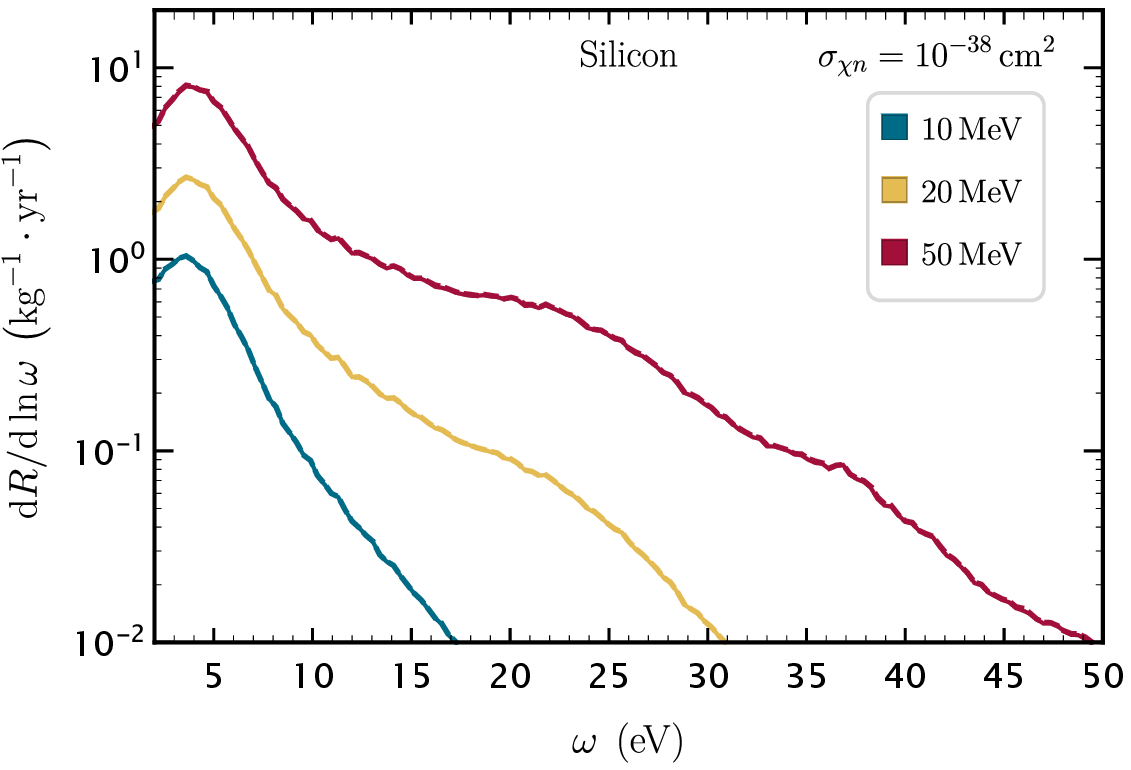}\hspace{0.7cm}\includegraphics[scale=0.745]{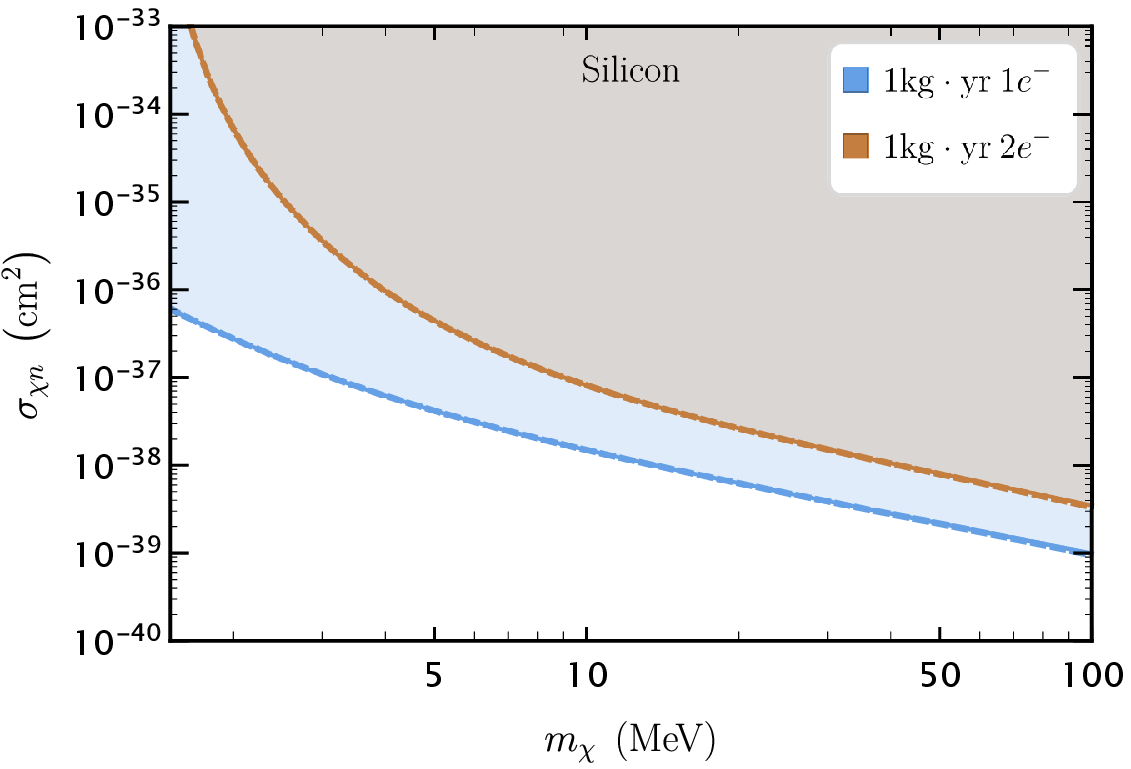}
\par\end{centering}
\caption{\textbf{\textit{Left}}: \label{fig:Comparison}The differential Migdal
electronic excitation event rates in bulk silicon for a reference
cross section $\sigma_{\chi n}=10^{-38}\,\mathrm{cm}^{2}$ and DM
masses $m_{\chi}=10\,\mathrm{MeV}$ (\textit{emerald}), $20\,\mathrm{MeV}$(\textit{orange})
and $50\,\mathrm{MeV}$ (\textit{maroon}), respectively; the \textit{solid}
lines and the \textit{dashed} lines are calculated using the phonon-mediated
approach and the bremsstrahlung-like approach, respectively. The results
from the two approaches coincide so well that they can hardly be distinguished
in the plot.\textbf{\textit{ Right}}: Sensitivities for the Migdal
effect at 90\% C.L. for a 1 kg$\cdot$yr silicon detector, based on
the single-electron (\textit{blue}) and the two-electron (\textit{orange})
ionization signal bins. The upper limits calculated using the phonon-mediated
approach~(\textit{solid}) and the bremsstrahlung-like approach~(\textit{dashed})
coincide throughout the whole DM mass range. See text for details.}
\end{figure}
\end{widetext}

It is interesting to compare the event rate~Eq.~(\ref{eq:eventRate-1})
with the one derived in the picture of the bremsstrahlung-like process
proposed in Ref.~\citep{Liang:2020ryg}, which is expressed as
\begin{eqnarray}
R & = & \frac{\rho_{\chi}}{m_{\chi}}\frac{2\,\alpha_{e}A^{2}\sigma_{\chi n}\,Z_{\mathrm{ion}}^{2}\,N_{T}}{3\,\Omega\,\mu_{\chi n}^{2}\,m_{N}^{2}}\,\int\mathrm{d^{3}}v\,\frac{f_{\chi}\left(\mathbf{v}\right)}{v}\int\frac{\mathrm{d}\omega}{\omega^{4}}\,\mathcal{F}\left(\omega\right)\,\int p{}_{N}^{3}\,\mathrm{d}p_{N}\times\varTheta\left[p_{N}v-\frac{p_{N}^{2}}{2\,\mu_{\chi N}}-\omega\right],\label{eq:brem_rate}
\end{eqnarray}
with $p_{N}$ and $\mu_{\chi N}=m_{N}\,m_{\chi}/\left(m_{N}+m_{\chi}\right)$
being momentum of the recoiled nucleus and the reduced mass of the
DM-nucleus pair, respectively. $\varTheta$ is the Heaviside step
function. In the sub-GeV mass regime, $\mu_{\chi N}\approx m_{\chi}$.
Since the integrand functions $\left\{ T_{n}\left(E\right)\right\} $
vanish if $E<0$, we set the lower limit of the integral to be $0$
in Eq.~(\ref{eq:eventRate-1}) for convenience. But note that $T_{0}\left(E\right)=\delta\left(E\right)$
also contributes to the event rate if $E>0$ in Eq.~(\ref{eq:eventRate-1}),
which corresponds to the process where an electron-hole pair is excited
without generating any phonons. 

In our computations, we take $\rho_{\chi}=0.3\,\mathrm{GeV/cm^{3}}$,
and the velocity distribution is approximated as a truncated Maxwellian
form in the Galactic rest frame, i.e\textit{.}, $f_{\chi}\left(\mathbf{v}\right)\propto\exp\left[-\left|\mathbf{v}+\mathbf{v}_{\mathrm{e}}\right|^{2}/v_{0}^{2}\right]\,\Theta\left(v_{\mathrm{esc}}-\left|\mathbf{v}+\mathbf{v}_{\mathrm{e}}\right|\right)$,
with the Earth's velocity $v_{\mathrm{e}}=230\,\mathrm{km/s}$, the
dispersion velocity $v_{0}=220\,\mathrm{km/s}$ and the Galactic escape
velocity $v_{\mathrm{esc}}=544\,\,\mathrm{km/s}$~\citep{Baxter:2021pqo}.
In the left panel of Fig.~\ref{fig:Comparison} shown is the comparison
between the Migdal event rates in silicon semiconductor target calculated
with the phonon-mediated approach and the bremsstrahlung-like method,
for DM masses $m_{\chi}=10\,\mathrm{MeV}$, $20\,\mathrm{MeV}$ and
$50\,\mathrm{MeV}$, respectively, for a benchmark cross section $\sigma_{\chi n}=10^{-38}\,\mathrm{cm}^{2}$.
It is observed that even in the small DM mass range~($m_{\chi}<50\,\mathrm{MeV}$),
where the impulse approximation is no longer expected to be reliable,
the bremsstrahlung-like narrative still well coincides with the result
calculated with the phonon-mediated approach. This can be partly understood
from a closer observation of $S\left(q,\,E\right)$: for a DM mass
below $10\,\mathrm{MeV}$ with a typical momentum transfer $q=m_{\chi}\cdot10^{-3}$$<10\,\mathrm{keV}$,
$S\left(q,\,E\right)$ is dominated by $T_{0}\left(E\right)=\delta\left(E\right)$,
and thus $\int^{qv-\frac{q^{2}}{2m_{\chi}}-\omega}S\left(q,\,E\right)\mathrm{d}E\simeq\varTheta\left(qv-\frac{q^{2}}{2m_{\chi}}-\omega\right)$,
in consistence with the bremsstrahlung-like expression in Eq.~(\ref{eq:brem_rate}).
For larger DM masses with a typical velocity $10^{-3}\,c$, $S\left(q,\,E\right)$
resembles a Gaussian form centered at $E_{R}=q^{2}/\left(2m_{N}\right)\ll q^{2}/\left(2m_{\chi}\right)\sim\mathcal{O}\left(qv-\frac{q^{2}}{2m_{\chi}}-\omega\right)$,
and again one has $\int^{qv-\frac{q^{2}}{2m_{\chi}}-\omega}S\left(q,\,E\right)\mathrm{d}E\simeq\varTheta\left(qv-\frac{q^{2}}{2m_{\chi}}-\omega\right)$.
This relation holds as long as the weight of $S\left(q,\,E\right)$
lies below $qv-\frac{q^{2}}{2m_{\chi}}-\omega$.

In the right panel of Fig.~\ref{fig:Comparison} we present the expected
90\% C.L. sensitivity of silicon target to cross section $\sigma_{\chi n}$
with $1$ kg$\cdot$yr of exposure, based on the phonon-mediated~(\textit{solid})
and bremsstrahlung-like~(\textit{dashed}) approaches, for a single-electron~(\textit{blue})
and a two-electron~(\textit{orange}) charge bins, respectively, under
the zero background assumption. 

\section{\label{sec:Conclusions}Summary and discussion}

In this paper we build up a phonon-mediated description of the Migdal
effect in semiconductor targets in the context of the solid state
QFT, in which the phonons, namely, the quantized collective vibrations
of the ions, rather than the on-shell ions, are used to describe the
Migdal excitation process. 

In order to ease the discussion, three major simplifications of the
problem are made: (1) we assume the solid target is a monatomic simple
crystal, possessing an approximately rotational symmetry; (2) we use
the zero-temperature QFT in formulating the multi-phonon scattering
event rates, rather than a more general finite temperature QFT framework;
(3) we take the incoherent approximation in calculating the multi-phonon
process. While the isotropy approximation in (1) is valid for the
diamond structure materials~(e.g., silicon and germanium), it may
result in uncertainty for some anisotropic materials~(e.g., sapphire).
The second assumption is sufficient for experiments operated at cryogenic
temperatures. In fact, it is straightforward to generalize the zero-temperature
formalism to the finite temperature one by simply replacing the propagators
of the zero-temperature case in Eq.~(\ref{eq:Amplitude_Electron_phonon})
with those in the finite temperature scenario. As for the third approximation,
note that for DM masses around an MeV, the typical momentum transfer
can be as small as $q\sim\mathcal{O}\left(1\right)\,\mathrm{keV}$,
which is comparable to the size of the 1BZ, and hence beyond the regime
of validity for the incoherent approximation. Further study for $m_{\chi}<1\,\mathrm{MeV}$
is needed.

Based on the formalism, we numerically calculate the Migdal excitation
event rates for the silicon semiconductor target. As expected, the
multi-phonon energy spectra are found to well converge to the Gaussian
form in the large $q$ limit, justifying the impulse approximation
used in the bremsstrahlung-like description. Although the behavior
of the phonon scattering function $S\left(q,\omega\right)$ differ
from that of a free nucleus in the low and intermediate scattering
energy region, the Migdal excitation rates calculated from the impulse
approximation are found to be well consistent with that obtained using
the phonon-mediated approach throughout the relevant DM mass range.
Finally, it is tempting to apply the phonon-mediated approach to the
probe of sub-MeV DM particles through the Migdal effect in novel narrow-gap
materials with band gaps of $\mathcal{O}\left(10\right)\mathrm{meV}$~(\textit{e.g.},
Dirac materials), where the picture of the free-recoiling nucleus
turns invalid altogether. We leave it for the future work.

\vspace{0.5cm}

\textbf{\textit{Note added.}} After this work was published, Kim Berghaus
suggested us that the contribution of $T_{0}$ was omitted in our
original numerical implementation of Eq.~(\ref{eq:eventRate-1}),
which as a consequence can remarkably suppress the calculated event
rates for a small $q$~(an upcoming paper~\citep{Berghaus:2022}
also discusses the Migdal effect in semiconductor detectors). After
$T_{0}$ term is included, it is found that the Migdal event rates
calculated from the phonon-mediated approach and the impulse approximation
coincide quite well even in the low DM mass range.

\appendix

\renewcommand{\theequation}{A.\arabic{equation}}
\begin{acknowledgments}
This work was partly supported by Science Challenge Project under
No.~TZ2016001, by the National Key R\&D Program of China under Grant
under No.~2017YFB0701502, and by National Natural Science Foundation
of China under No.~11625415. C.M. was supported by the NSFC under
Grants No.~12005012, No.~11947202, and No.~U1930402, and by the
China Postdoctoral Science Foundation under Grants No.~2020T130047
and No.~2019M660016.
\end{acknowledgments}

\section{\label{sec:Phonons_appendix}Phonons in the quantum field theory}

For easy reading, we provide an elementary introduction to relevant
theoretical background to the main text in this appendix, including
the treatment of the phonons and electrons, as well as their interactions
in the context of the QFT. Part of the material can be found in Ref.~\citep{book:17984}.

\subsection{Quantization of vibrations in solids}

For simplicity here we only consider the case of the monatomic simple
lattices. The dynamics of the crystal vibration is described with
the following equation,
\begin{eqnarray}
m_{N}\ddot{\mathbf{u}}_{\boldsymbol{\ell}}+\sum_{\boldsymbol{\ell}'}\Phi\left(\boldsymbol{\ell}-\boldsymbol{\ell}'\right)\,\mathbf{u}_{\boldsymbol{\ell}'} & = & 0,
\end{eqnarray}
where $m_{N}$ is the nucleus mass, $\mathbf{u}_{\boldsymbol{\ell}}$
is the displacement of the nucleus at lattice site $\boldsymbol{\ell}$,
and the force strength matrix elements are explicitly expressed as
follows, 
\begin{eqnarray}
\Phi\left(\boldsymbol{\ell}-\boldsymbol{\ell}'\right)_{\sigma\sigma'} & = & \left.\frac{\partial^{2}U}{\partial u_{\boldsymbol{\ell}\sigma}\,\partial u_{\boldsymbol{\ell}'\sigma'}}\right|_{\mathbf{u}=0},
\end{eqnarray}
with $U$ being the potential between the nuclei at sites $\boldsymbol{\ell}$
and $\boldsymbol{\ell}'$, and $\sigma,\,\sigma'=\left\{ x,\,y,\,z\right\} $
denoting the three space directions. The Fourier transform of the
force strength matrix is called the dynamical matrix $V\left(\mathbf{k}\right)$,
\textit{i.e.}, 
\begin{eqnarray}
V\left(\mathbf{k}\right) & = & \sum_{\ell}\Phi\left(\boldsymbol{\ell}\right)e^{-i\mathbf{k}\cdot\boldsymbol{\ell}},
\end{eqnarray}
which is real symmetric matrix for the Bravais lattice at each wave-vector
$\mathbf{k}$ in the 1BZ, and thus can be diagonalized with an orthonormal
basis set of vectors $\left\{ \mathbf{\boldsymbol{\epsilon}}_{\mathbf{k},1},\,\mathbf{\boldsymbol{\epsilon}}_{\mathbf{k},2},\,\mathbf{\boldsymbol{\epsilon}}_{\mathbf{k},3}\right\} $,
such that
\begin{eqnarray}
\mathbf{\boldsymbol{\epsilon}}_{\mathbf{k},\alpha}\cdot\mathbf{\boldsymbol{\epsilon}}_{\mathbf{k},\alpha'} & = & \delta_{\alpha,\alpha'}.
\end{eqnarray}
and 
\begin{eqnarray}
\mathbf{\boldsymbol{\epsilon}}_{\mathbf{k},\alpha'}^{T}V\left(\mathbf{k}\right)\mathbf{\boldsymbol{\epsilon}}_{\mathbf{k},\alpha'} & = & \omega_{\mathbf{k},\alpha}^{2}\delta_{\alpha,\alpha'},
\end{eqnarray}
with $\omega\left(\mathbf{k},\alpha\right)$ being corresponding eigenfrequency
of the mode $\left(\mathbf{k},\,\alpha\right)$. Besides, since $\Phi\left(\boldsymbol{\ell}\right)=\Phi\left(-\boldsymbol{\ell}\right)$
for Bravais lattice, it is straightforward to see that $V\left(\mathbf{k}\right)=V\left(-\mathbf{k}\right)$,
$\mathbf{\boldsymbol{\epsilon}}_{\mathbf{k},\alpha}=\mathbf{\boldsymbol{\epsilon}}_{-\mathbf{k},\alpha}$
and $\omega_{\mathbf{k},\alpha}=\omega_{-\mathbf{k},\alpha}$. With
these preparations, one first substitutes the displacements with eigen-vibration
modes:
\begin{eqnarray}
\mathbf{u}_{\boldsymbol{\ell}} & = & \sum_{\mathbf{k}\in1\mathrm{BZ}}\,\sum_{\alpha=1}^{3}\frac{\mathbf{\boldsymbol{\epsilon}}_{\mathbf{k},\alpha}\,e^{i\mathbf{k}\cdot\boldsymbol{\ell}}}{\sqrt{N\,m_{N}}}\,Q_{\mathbf{k},\alpha},
\end{eqnarray}
where $Q_{\mathbf{k},\alpha}$ encodes the vibration amplitude for
the mode $\left(\mathbf{k},\,\alpha\right)$, and $N$ is the number
of the unit cells in the material, and then obtains the Lagrangian
of vibration system,
\begin{eqnarray}
L & = & \frac{1}{2}\left(\sum_{\boldsymbol{\ell}}m_{N}\,\mathbf{\dot{u}}_{\boldsymbol{\ell}}\cdot\mathbf{\dot{u}}_{\boldsymbol{\ell}}-\sum_{\boldsymbol{\ell},\boldsymbol{\ell}'}\mathbf{u}_{\boldsymbol{\ell}}^{T}\,\Phi\left(\boldsymbol{\ell}-\boldsymbol{\ell}'\right)\mathbf{u}_{\boldsymbol{\ell}'}\right)\nonumber \\
 & = & \frac{1}{2}\sum_{\mathbf{k},\alpha}\left(\dot{Q}_{-\mathbf{k},\alpha}\,\dot{Q}_{\mathbf{k},\alpha}-\omega_{\mathbf{k},\alpha}^{2}Q_{-\mathbf{k},\alpha}\,Q_{\mathbf{k},\alpha}\right),
\end{eqnarray}
and the equations of motion
\begin{eqnarray}
\ddot{Q}_{\mathbf{k},\alpha}+\omega_{\mathbf{k},\alpha}^{2}Q_{\mathbf{k},\alpha} & = & 0.
\end{eqnarray}
One then follows the conventional quantization procedures to give
pairs of the canonical position and momentum operators
\begin{eqnarray}
\hat{Q}_{\mathbf{k},\alpha}\left(t\right) & = & \frac{1}{\sqrt{2\,\omega_{\mathbf{k},\alpha}}}\left(\hat{a}_{\mathbf{k},\alpha}e^{-i\omega_{\mathbf{k},\alpha}t}+\hat{a}_{-\mathbf{k},\alpha}^{\dagger}e^{i\omega_{\mathbf{k},\alpha}t}\right),\nonumber \\
\hat{P}_{\mathbf{k},\alpha}\left(t\right) & = & -i\sqrt{\frac{\omega_{\mathbf{k},\alpha}}{2}}\left(\hat{a}_{-\mathbf{k},\alpha}e^{-i\omega_{\mathbf{k},\alpha}t}-\hat{a}_{\mathbf{k},\alpha}^{\dagger}e^{i\omega_{\mathbf{k},\alpha}t}\right),
\end{eqnarray}
that satisfy the equal time commutation relation $\left[\hat{Q}_{\mathbf{k},\alpha}\left(t\right),\,\hat{P}_{\mathbf{k}',\alpha'}\left(t\right)\right]=i\,\delta_{\mathbf{k},\mathbf{k}'}\,\delta_{\alpha,\alpha'}$,
and $\left[\hat{Q}_{\mathbf{k},\alpha}\left(t\right),\,\hat{Q}_{\mathbf{k}',\alpha'}\left(t\right)\right]=\left[\hat{P}_{\mathbf{k},\alpha}\left(t\right),\,\hat{P}_{\mathbf{k}',\alpha'}\left(t\right)\right]=0$,
from the commutation relation $\left[\hat{a}_{\mathbf{k},\alpha},\,\hat{a}_{\mathbf{k}',\alpha'}^{\dagger}\right]=\delta_{\mathbf{k},\mathbf{k}'}\,\delta_{\alpha,\alpha'}$
and $\left[\hat{a}_{\mathbf{k},\alpha},\,\hat{a}_{\mathbf{k}',\alpha'}\right]=\left[\hat{a}_{\mathbf{k},\alpha}^{\dagger},\,\hat{a}_{\mathbf{k}',\alpha'}^{\dagger}\right]=0$,
and vice versa. In the context of the path integral quantization,
the action is written as 
\begin{eqnarray}
\int L_{\mathrm{phonon}}\mathrm{\,d}t & = & \int\frac{1}{2}\sum_{\mathbf{k},\alpha}\left(\dot{Q}_{-\mathbf{k},\alpha}\,\dot{Q}_{\mathbf{k},\alpha}-\omega_{\mathbf{k},\alpha}^{2}Q_{-\mathbf{k},\alpha}\,Q_{\mathbf{k},\alpha}\right)\mathrm{\,d}t\nonumber \\
 & = & \int\frac{1}{2}\sum_{\mathbf{k},\alpha}Q_{-\mathbf{k},\alpha}\overrightarrow{\left(-\partial_{t}^{2}-\omega_{\mathbf{k},\alpha}^{2}\right)}\,Q_{\mathbf{k},\alpha}\mathrm{\,d}t,
\end{eqnarray}
from which one constructs the free phonon propagator
\begin{eqnarray}
iD_{\mathbf{k},\alpha}\left(t-t'\right) & = & \int\frac{i}{\omega^{2}-\omega_{\mathbf{k},\alpha}^{2}+i0^{+}}\frac{e^{-i\omega\left(t-t'\right)}\mathrm{d}\omega}{2\pi}
\end{eqnarray}
satisfying $\overrightarrow{\left(-\partial_{t}^{2}-\omega_{\mathbf{k},\alpha}^{2}\right)}\,D_{\mathbf{k},\alpha}\left(t-t'\right)=\delta\left(t-t'\right)$.

\subsection{Quantization of electrons in solids}

One can construct the path integral formalism for electrons in solids
in a similar fashion, except for the anti-commutation nature of the
Grassmann algebra. Here we summarize some important results. 

The action of the electron field can be drawn from the Schrödinger
equation as the following:
\begin{eqnarray}
\int\mathcal{L}_{\mathrm{electron}}\mathrm{\,d}^{4}x & = & \int\mathrm{\,d}^{4}x\left(i\psi_{e}^{*}\dot{\psi_{e}}-\frac{\nabla\psi_{e}^{*}\nabla\psi_{e}}{2m_{e}}-V\psi_{e}^{*}\psi_{e}\right)\nonumber \\
 & = & \int\mathrm{\,d}^{4}x\,\psi_{e}^{*}\overrightarrow{\left[i\frac{\partial}{\partial t}-\left(-\frac{\nabla^{2}}{2m_{e}}+V\right)\right]}\psi_{e},
\end{eqnarray}
where $-\nabla^{2}\psi_{e}/2m_{e}+V$ is Hamiltonian for a single
electron, with $\left\{ u_{i}\left(\mathbf{x}\right)\right\} $ and
$\left\{ \varepsilon_{i}\right\} $ being its eigenwavefunctions and
corresponding energies, respectively. Thus one obtains the electron
propagator
\begin{eqnarray}
S\left(x,\,x'\right) & = & \sum_{i}\int\frac{u_{i}\left(\mathbf{x}\right)u_{i}^{*}\left(\mathbf{x}'\right)}{\omega-\varepsilon_{i}+i0^{+}}\frac{e^{-i\omega\left(t_{x}-t_{x'}\right)}\mathrm{d}\omega}{2\pi}.
\end{eqnarray}
\vspace{0.3cm}

\subsection{\label{subsec:DM-phonon}DM-phonon interaction}

\begin{figure}[h]
\begin{centering}
\includegraphics[scale=0.85]{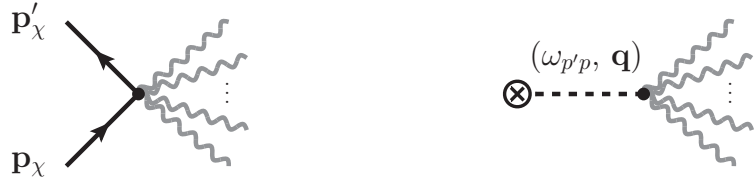}
\par\end{centering}
\caption{\label{fig:externalField}The effects of the incident DM particle
on the target material can be regarded as an external field. }
\end{figure}

The coupling term in the action between the DM particle and nuclei
in solids can be directly written as 
\begin{eqnarray}
\mathcal{V}_{\chi N} & = & -\int\mathrm{\,d}^{4}x\,\mathrm{\,d}^{4}x'\,\psi_{\chi}^{*}\left(x\right)\psi_{\chi}\left(x\right)\,V_{\chi N}\left(x-x'\right)\,\sum_{\boldsymbol{\ell}}\delta^{3}\left(\mathbf{x}'-\boldsymbol{\ell}-\mathbf{u}_{\boldsymbol{\ell}}\right),
\end{eqnarray}
where the interaction is instantaneous such that $V_{\chi N}\left(x-x'\right)=V_{\chi N}\left(\mathbf{x}-\mathbf{x}'\right)\,\delta\left(t-t'\right)$.
Since one is only interested in the target material that hosts the
phonons and electrons, it is convenient to integrate out the DM component
and regard it as an external field. To be specific, one draws DM external
field from the amplitude of the scattering process $\chi\left(\mathbf{p}_{\chi}\right)+$target$\rightarrow$$\chi\left(\mathbf{p}'_{\chi}\right)+$target~(excited)~(see
Fig.~\ref{fig:externalField} for illustration) and obtains the effective
Lagrangian
\begin{eqnarray}
L_{\chi\mathrm{phonon}} & = & -\frac{V_{\chi N}\left(\mathbf{q}\right)e^{i\omega_{p'p}t}}{V}\left[\int\mathrm{d}^{3}x'\,e^{-i\mathbf{q}\cdot\mathbf{x}'}\,\sum_{\boldsymbol{\ell}}\delta^{3}\left(\mathbf{x}'-\boldsymbol{\ell}-\mathbf{u}_{\boldsymbol{\ell}}\left(t\right)\right)\right]\nonumber \\
 & = & -\frac{V_{\chi N}\left(\mathbf{q}\right)e^{i\omega_{p'p}t}}{V}\sum_{\boldsymbol{\ell}}e^{-i\mathbf{q}\cdot\left(\boldsymbol{\ell}+\mathbf{u}_{\boldsymbol{\ell}}\left(t\right)\right)}\nonumber \\
 & = & -\frac{V_{\chi N}\left(\mathbf{q}\right)e^{i\omega_{p'p}t}}{V}\sum_{s=0}^{+\infty}\frac{1}{s!}\sum_{\left\{ \mathbf{p}_{j},\,\beta_{j}\,;s\right\} }\sum_{\boldsymbol{\ell}}e^{i\left(\sum_{j}\mathbf{p}_{j}-\mathbf{q}\right)\cdot\boldsymbol{\ell}}\left(\frac{-i\mathbf{q}\cdot\mathbf{\boldsymbol{\epsilon}}_{\mathbf{p}_{1},\beta_{1}}}{\sqrt{N\,m_{N}}}\right)\cdots\left(\frac{-i\mathbf{q}\cdot\mathbf{\boldsymbol{\epsilon}}_{\mathbf{p}_{s},\beta_{s}}}{\sqrt{N\,m_{N}}}\right)Q_{\mathbf{p}_{1},\beta_{1}}\cdots Q_{\mathbf{p}_{s},\beta_{s}}\nonumber \\
 & = & -\frac{V_{\chi N}\left(\mathbf{q}\right)e^{i\omega_{p'p}t}}{V}\sum_{s=0}^{+\infty}\frac{1}{s!}\sum_{\left\{ \mathbf{p}_{j},\,\beta_{j}\,;s\right\} }N\,\sum_{\mathbf{G}}\delta_{\sum_{j}\mathbf{p}_{j}-\mathbf{q},\,\mathbf{G}}\left(\frac{-i\mathbf{q}\cdot\mathbf{\boldsymbol{\epsilon}}_{\mathbf{p}_{1},\beta_{1}}}{\sqrt{N\,m_{N}}}\right)\cdots\left(\frac{-i\mathbf{q}\cdot\mathbf{\boldsymbol{\epsilon}}_{\mathbf{p}_{s},\beta_{s}}}{\sqrt{N\,m_{N}}}\right)Q_{\mathbf{p}_{1},\beta_{1}}\cdots Q_{\mathbf{p}_{s},\beta_{s}},\nonumber \\
\end{eqnarray}
where $\mathbf{q}=\mathbf{p}'_{\chi}-\mathbf{p}_{\chi}$, and $\omega_{p'p}=\left|\mathbf{p}'_{\chi}\right|^{2}/2\,m_{\chi}-\left|\mathbf{p}{}_{\chi}\right|^{2}/2\,m_{\chi}$,
and $V_{\chi N}\left(\mathbf{q}\right)$ is the Fourier transform
of the DM-nucleus contact interaction $V_{\chi N}\left(\mathbf{x}\right)$.
It should be noted that in above discussion we adopt the discrete
momentum convention. 

Based on above discussion, one can derive the \textit{LSZ reduction
formula} for the multi-phonon scattering process. For example, the
\textit{S}-matrix for an $n$-phonon scattering process subject to
a DM external field can be expressed as
\begin{eqnarray}
 &  & -i\int\mathrm{d}t\,\frac{V_{\chi N}\left(\mathbf{q}\right)e^{i\omega_{p'p}t}}{V}\sum_{s=0}^{+\infty}\frac{1}{s!}\,\sum_{\left\{ \mathbf{p}_{j},\,\beta_{j}\,;s\right\} }N\,\sum_{\mathbf{G}}\delta_{\sum_{j}\mathbf{p}_{j}-\mathbf{q},\,\mathbf{G}}\,\prod_{i=1}^{n}\int\mathrm{d}t_{i}\,e^{i\omega_{\mathbf{k}_{i},\alpha_{i}}t_{i}}\frac{i}{\sqrt{2\,\omega_{\mathbf{k}_{i},\alpha_{i}}}}\overrightarrow{\left(\frac{\partial^{2}}{\partial t_{i}^{2}}+\omega_{\mathbf{k}_{i},\alpha_{i}}^{2}\right)}\,\nonumber \\
 &  & \times\left.\overrightarrow{\frac{\delta}{\delta iJ_{\mathbf{k}_{1},\alpha_{1}}}}\cdots\overrightarrow{\frac{\delta}{\delta iJ_{\mathbf{k}_{n},\alpha_{n}}}}\left[e^{\sum_{\mathbf{k},\alpha}iJ_{\mathbf{k},\alpha}\frac{iD_{\mathbf{k},\alpha}}{2}iJ_{-\mathbf{k},\alpha}}\right]\overleftarrow{\frac{\delta}{\delta iJ_{\mathbf{p}_{1},\beta_{1}}}}\cdots\overleftarrow{\frac{\delta}{\delta iJ_{\mathbf{p}_{s},\beta_{s}}}}\right|_{J=0}\left(\frac{-i\mathbf{q}\cdot\mathbf{\boldsymbol{\epsilon}}_{\mathbf{p}_{1},\beta_{1}}}{\sqrt{N\,m_{N}}}\right)\cdots\left(\frac{-i\mathbf{q}\cdot\mathbf{\boldsymbol{\epsilon}}_{\mathbf{p}_{s},\beta_{s}}}{\sqrt{N\,m_{N}}}\right),\nonumber \\
\label{eq:LSZ}
\end{eqnarray}
where $\left\{ \mathbf{k}_{i},\alpha_{i}\right\} \,\left(i=1,\,2,\cdots,\,n\right)$
label the $n$-phonon final states. 

\begin{figure}
\begin{centering}
\includegraphics{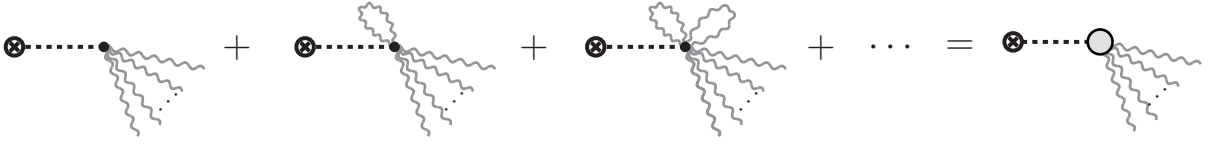}
\par\end{centering}
\caption{\label{fig:DebyeWaller}The diagram for the $n$-phonon scattering
process containing full contributions of phonon loops, which generates
the Debye-Waller factor denoted as the gray blob on the right-hand-side
of the equation. See text for details.}

\end{figure}
\begin{figure}
\begin{centering}
\includegraphics[scale=0.5]{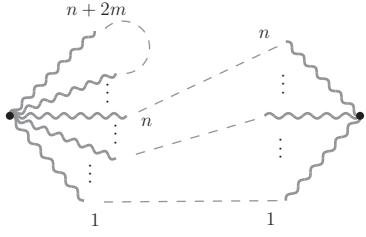}
\par\end{centering}
\caption{\label{fig:SymmetryFactor} Illustration of possible ways in which
$n$ phonon external lines connect to the $n$ internal ones while
the left $2m$ internal members self-connect with each other to form
$m$ loops. See text for details.}
\end{figure}

Now one is able to investigate the amplitude for the full $n$-phonon
scattering process shown in Fig.~\ref{fig:DebyeWaller}, which contains
not only the tree level piece, but also higher orders of the phonon
loop. Note that a self-closed propagator for mode $\left(\mathbf{k},\alpha\right)$
reads as 
\begin{eqnarray}
iD_{\mathbf{k},\alpha}\left(t^{+}-t\right) & = & \int_{-\infty}^{+\infty}\frac{i}{\omega^{2}-\omega_{\mathbf{k},\alpha}^{2}+i0^{+}}\frac{e^{-i0^{+}\omega}\mathrm{d}\omega}{2\pi}\nonumber \\
 & = & \frac{1}{2\,\omega_{\mathbf{k},\alpha}},
\end{eqnarray}
as well as a companion factor $\left(\frac{-i\mathbf{q}\cdot\mathbf{\boldsymbol{\epsilon}}_{\mathbf{k},\alpha}}{\sqrt{N\,m_{N}}}\right)\left(\frac{-i\mathbf{q}\cdot\mathbf{\boldsymbol{\epsilon}}_{\mathbf{k},\alpha}}{\sqrt{N\,m_{N}}}\right)$.
Thus every loop in diagram corresponds to a factor $-\sum_{\mathbf{k},\alpha}\frac{\left|\mathbf{q}\cdot\mathbf{\boldsymbol{\epsilon}}_{\mathbf{k},\alpha}\right|^{2}}{2N\,m_{N}\omega_{\mathbf{k},\alpha}}$.
On the other hand, a specific external leg $\left(\mathbf{k}_{i},\alpha_{i}\right)$
corresponds to 
\begin{eqnarray}
\int\mathrm{d}t_{i}\,e^{i\omega_{\mathbf{k}_{i},\alpha_{i}}t_{i}}\frac{i}{\sqrt{2\,\omega_{\mathbf{k}_{i},\alpha_{i}}}}\overrightarrow{\left(\frac{\partial^{2}}{\partial t_{i}^{2}}+\omega_{\mathbf{k}_{i},\alpha_{i}}^{2}\right)}\,iD_{\mathbf{k}_{i},\alpha_{i}}\left(t_{i}-t\right) & = & \frac{e^{i\omega_{\mathbf{k}_{i},\alpha_{i}}t}}{\sqrt{2\,\omega_{\mathbf{k}_{i},\alpha_{i}}}}.
\end{eqnarray}
Besides, the effect of the symmetry factor should also be taken into
account. For example, one considers the $n$-phonon process containing
$m$ self-interacting loops in Fig.~\ref{fig:DebyeWaller}. Determining
the number of the contractions in Eq.~(\ref{eq:LSZ}) is equivalent
to enumerating all possible ways the internal phonon lines interconnect
among themselves, which is illustrated in Fig.~\ref{fig:SymmetryFactor}.
It is not difficult to verify that the overall constant that encodes
the symmetry effect is equal to $\frac{\left(2m\right)!}{m!\,2^{m}}C_{n+2m}^{2m}\frac{n!}{\left(n+2m\right)!}=\frac{1}{m!\,2^{m}}$,
where $C_{n+2m}^{2m}$counts which $2m$ lines self-connect among
all $n+2m$ lines from the left in Fig.~\ref{fig:SymmetryFactor},
$\frac{\left(2m\right)!}{m!\,2^{m}}$ is the number of the ways these
specific $2m$ lines connect with each other, $n!$ describes the
interchange of the external legs at the right hand in Fig.~\ref{fig:SymmetryFactor},
and $\frac{1}{\left(n+2m\right)!}$ corresponds to the factor $\frac{1}{s!}$
in Eq.~(\ref{eq:LSZ}).

Putting all these pieces together, the sum of all diagrams on the
left-hand-side in Fig.~\ref{fig:DebyeWaller} can be expressed as
\begin{eqnarray}
 &  & \left(-i\right)\frac{V_{\chi N}\left(\mathbf{q}\right)}{V}N\,\sum_{\mathbf{G}}\delta_{\sum_{i}\mathbf{k}_{i}+\mathbf{q},\,\mathbf{G}}\,\prod_{i=1}^{n}\left(\frac{-i\mathbf{q}\cdot\mathbf{\boldsymbol{\epsilon}}_{\mathbf{k}_{i},\alpha_{i}}}{\sqrt{2\,N\,m_{N}\omega_{\mathbf{k}_{i},\alpha_{i}}}}\right)\sum_{m=0}^{+\infty}\frac{1}{m!}\,\left(-\sum_{\mathbf{k},\alpha}\frac{\left|\mathbf{q}\cdot\mathbf{\boldsymbol{\epsilon}}_{\mathbf{k},\alpha}\right|^{2}}{4N\,m_{N}\omega_{\mathbf{k},\alpha}}\right)^{m}\int e^{i\left(\sum_{i}\omega_{\mathbf{k}_{i},\alpha_{i}}+\omega_{p'p}\right)t}\mathrm{d}t\nonumber \\
 & = & \left(-i\right)\frac{V_{\chi N}\left(\mathbf{q}\right)}{V}N\,\sum_{\mathbf{G}}\delta_{\sum_{i}\mathbf{k}_{i}+\mathbf{q},\,\mathbf{G}}\,\prod_{i=1}^{n}\left(\frac{-i\mathbf{q}\cdot\mathbf{\boldsymbol{\epsilon}}_{\mathbf{k}_{i},\alpha_{i}}}{\sqrt{2\,N\,m_{N}\omega_{\mathbf{k}_{i},\alpha_{i}}}}\right)e^{-\sum_{\mathbf{k},\alpha}\frac{\left|\mathbf{q}\cdot\mathbf{\boldsymbol{\epsilon}}_{\mathbf{k},\alpha}\right|^{2}}{4Nm_{N}\omega_{\mathbf{k},\alpha}}}\,2\pi\delta\left(\sum_{i}\omega_{\mathbf{k}_{i},\alpha_{i}}+\omega_{p'p}\right),
\end{eqnarray}
where $e^{-\sum_{\mathbf{k},\alpha}\frac{\left|\mathbf{q}\cdot\mathbf{\boldsymbol{\epsilon}}_{\mathbf{k},\alpha}\right|^{2}}{4Nm_{N}\omega_{\mathbf{k},\alpha}}}$
is no other than the Debye-Waller factor at the zero-temperature,
which is represented with the gray blob on the right-hand-side of
Fig.~\ref{fig:DebyeWaller}. In the derivation we interchange $\mathbf{\boldsymbol{\epsilon}}_{\mathbf{k},\alpha}$
and $\mathbf{\boldsymbol{\epsilon}}_{-\mathbf{k},\alpha}$ whenever
necessary. From above discussion we can see one benefit of the path
integral approach: one no longer has to resort to the cumbersome operator
commutator algebra to obtain the Debye-Waller factor. Propagators
do the job.

\subsection{\label{subsec:Phonon-electron}Electron-phonon interaction}

The interaction between the ions and electrons can be directly written
as 
\begin{eqnarray}
V_{\mathrm{ion}-e}\left(\mathbf{x}\right) & = & \sum_{\boldsymbol{\ell}}U_{e}\left(\mathbf{x}-\boldsymbol{\ell}-\mathbf{u}_{\boldsymbol{\ell}}\right)\nonumber \\
 & = & \sum_{\boldsymbol{\ell}}U_{e}\left(\mathbf{x}-\boldsymbol{\ell}\right)+\sum_{\boldsymbol{\ell}}\left(-\mathbf{u}_{\boldsymbol{\ell}}\right)\cdot\nabla U_{e}\left(\mathbf{x}-\boldsymbol{\ell}\right),
\end{eqnarray}
where $U_{e}\left(\mathbf{x}-\boldsymbol{\ell}\right)=-Z_{\mathrm{ion}}\alpha_{e}/\left|\mathbf{x}-\boldsymbol{\ell}\right|$
is Coulomb potential between the ion located at $\boldsymbol{\ell}$
and an electron at position $\mathbf{x}$. Thus, the electron-phonon
interaction Lagrangian term is written as
\begin{eqnarray}
\mathcal{L}_{\mathrm{phonon}-e} & = & \psi_{e}^{*}\left(x\right)\psi_{e}\left(x\right)\sum_{\boldsymbol{\ell}}\mathbf{u}_{\boldsymbol{\ell}}\cdot\nabla U_{e}\left(\mathbf{x}-\boldsymbol{\ell}\right)\nonumber \\
 & = & \psi_{e}^{*}\left(x\right)\psi_{e}\left(x\right)\sum_{\mathbf{k},\alpha}\mathbf{\boldsymbol{\epsilon}}_{\mathbf{k},\alpha}\cdot\nabla\left[\sum_{\boldsymbol{\ell}}U_{e}\left(\mathbf{x}-\boldsymbol{\ell}\right)e^{i\mathbf{k}\cdot\boldsymbol{\ell}}\right]\frac{Q_{\mathbf{k},\alpha}}{\sqrt{N\,m_{N}}}\nonumber \\
 & = & -\left(\frac{NZ_{\mathrm{ion}}}{V}\right)\psi_{e}^{*}\left(x\right)\psi_{e}\left(x\right)\,\sum_{\mathbf{k},\alpha}\sum_{\mathbf{G}}i\,\mathbf{\boldsymbol{\epsilon}}_{\mathbf{k},\alpha}\cdot\left(\mathbf{k}+\mathbf{G}\right)\,v\left(\mathbf{k}+\mathbf{G}\right)\,e^{i\mathbf{\left(\mathbf{k}+\mathbf{G}\right)}\cdot\mathbf{x}}\,\frac{Q_{\mathbf{k},\alpha}}{\sqrt{N\,m_{N}}},
\end{eqnarray}
with $v\left(\mathbf{k}+\mathbf{G}\right)=4\pi\alpha_{e}/\left|\mathbf{k}+\mathbf{G}\right|^{2}$.

\subsection{\label{subsec:Feynman-rules}Feynman rules}

Based upon above preparation, the Feynman rules describing the processes
involving multi-phonon and electron-phonon interactions in momentum
space can be summarized as follows.
\begin{itemize}
\item In discussion the vertex of the DM particle is replaced with an external
field as shown in Fig.~\ref{fig:externalField}. Such an external
source corresponds $\left(-i\right)V_{\chi N}\left(\mathbf{q}\right)/V$.
\item A gray blob corresponds to the factor $e^{-W\left(\mathbf{q}\right)}=e^{-\sum_{\mathbf{k},\alpha}\frac{\left|\mathbf{q}\cdot\mathbf{\boldsymbol{\epsilon}}_{\mathbf{k},\alpha}\right|^{2}}{4Nm_{N}\omega_{\mathbf{k},\alpha}}}$.
\item Each blob also contributes the energy-momentum conservation condition
presented as discrete delta functions $N\,\sum_{\mathbf{G}}\delta_{\sum_{i}\mathbf{p}_{i},\,\mathbf{G}}\,2\pi\delta\left(\sum\varepsilon_{i}\right)$,
with $\left\{ \mathbf{p}_{i}\right\} $~($\left\{ \varepsilon_{i}\right\} $)
being the momenta~(energies) flowing out of the blob.
\item A phonon external leg representing the final state $\left(\mathbf{k}_{i},\alpha_{i}\right)$
from the blob corresponds to $\frac{-i\mathbf{q}\cdot\mathbf{\boldsymbol{\epsilon}}_{\mathbf{k}_{i},\alpha_{i}}}{\sqrt{2Nm_{N}\omega_{\mathbf{k}_{i},\alpha_{i}}}}$;
a phonon internal line with mode $\left(\mathbf{k},\alpha\right)$
connecting the blob comes with a factor $\frac{-i\mathbf{q}\cdot\mathbf{\boldsymbol{\epsilon}}_{\mathbf{k},\alpha}}{\sqrt{Nm_{N}}}$.
\item Each phonon internal line contributes a factor $1/2\pi$.
\item Vertex that contains both the incoming and outgoing states ($\ket{j}$
and $\ket{i}$) of the electrons in solids contributes a factor $\left(2\pi\right)\braket{i|e^{i\mathbf{p}\cdot\hat{\mathbf{x}}}|j}$
and the energy conservation condition $\delta\left(\sum_{i}\varepsilon_{i}\right)$,
where $\mathbf{p}$ is the net momentum sinking into the vertex.
\item A phonon-electron vertex is read as
\[
\left(-NZ_{\mathrm{ion}}\right)\sum_{\mathbf{G}}\frac{i\,\mathbf{\boldsymbol{\epsilon}}_{\mathbf{k},\alpha}\cdot\left(\mathbf{k}+\mathbf{G}\right)}{\sqrt{Nm_{N}}}\frac{v\left(\mathbf{k}+\mathbf{G}\right)}{V}\braket{i|e^{i\left(\mathbf{k}+\mathbf{G}\right)\cdot\hat{\mathbf{x}}}|j}.
\]
\item A phonon internal line with one end connecting a phonon blob and the
other connecting an electron vertex corresponds to the sum over propagators
of all modes $\left\{ \mathbf{k},\alpha\right\} $, i.e.,
\[
\sum_{\mathbf{k},\alpha}\left(\frac{-i\mathbf{q}\cdot\mathbf{\boldsymbol{\epsilon}}_{\mathbf{k},\alpha}}{\sqrt{Nm_{N}}}\right)\left(\frac{i}{\omega^{2}-\omega_{\mathbf{k},\alpha}^{2}+i0^{+}}\right)\left[\sum_{\mathbf{G}}\frac{i\,\mathbf{\boldsymbol{\epsilon}}_{\mathbf{k},\alpha}\cdot\left(\mathbf{k}+\mathbf{G}\right)}{\sqrt{Nm_{N}}}\frac{\left(-NZ_{\mathrm{ion}}\right)}{V}v\left(\mathbf{k}+\mathbf{G}\right)\braket{i|e^{i\left(\mathbf{k}+\mathbf{G}\right)\cdot\hat{\mathbf{x}}}|j}\right].
\]
\end{itemize}

\subsection{\label{subsec:RPA}Random phase approximation}

A short review of the \textit{random phase approximation}~(RPA) has
been provided in Ref.~\citep{Liang:2020ryg}, so here we only summarize
some results relevant for our present discussion. Within the framework
of the RPA, the Lindhard dielectric function for homogeneous electron
gas~(HEG) is expressed as 
\begin{eqnarray}
\epsilon\left(\mathbf{q},\omega\right) & = & 1-\frac{v\left(\mathbf{q}\right)}{V}\sum_{i,j}\frac{\left|\braket{i|e^{i\mathbf{q}\cdot\hat{\mathbf{x}}}|j}\right|^{2}}{\varepsilon_{i}-\varepsilon_{j}-\omega-i0^{+}}\left(n_{i}-n_{j}\right),\label{eq:RPA-dielectric-fucnition0}
\end{eqnarray}
where $n_{i}$~($n_{j}$) denotes the occupation number of the state
$\ket{i}$~($\ket{j}$), with $\varepsilon_{i}$~($\varepsilon_{j}$)
being corresponding eigenenergy. In crystalline structure the translational
symmetry for space-time reduces to that for the periodic crystal lattice.
The momentum transfer $\mathbf{q}$ in Eq.~(\ref{eq:RPA-dielectric-fucnition0})
is expressed uniquely as the sum of a reciprocal lattice vector $\mathbf{G}$,
and corresponding reduced momentum $\mathbf{\mathbf{k}}$ confined
in the 1BZ, i.e., $\mathbf{q}=\mathbf{k}+\mathbf{G}$. In this case,
the microscopic dielectric matrix 
\begin{eqnarray}
\widetilde{\epsilon}_{\mathbf{G},\mathbf{G}'}\left(\mathbf{\mathbf{k}},\omega\right) & = & \delta_{\mathbf{G},\mathbf{G}'}-\frac{1}{V}\frac{4\pi\alpha_{e}}{\left|\mathbf{\mathbf{k}}+\mathbf{G}\right|\left|\mathbf{\mathbf{k}}+\mathbf{G}'\right|}\sum_{i,j}\frac{\braket{i|e^{i\left(\mathbf{\mathbf{\mathbf{\mathbf{k}}+\mathbf{G}}}'\right)\cdot\hat{\mathbf{x}}}|j}\braket{j|e^{-i\left(\mathbf{\mathbf{k+\mathbf{G}}}\right)\cdot\hat{\mathbf{x}}}|i}}{\varepsilon_{i}-\varepsilon_{j}-\omega-i0^{+}}\left(n_{i}-n_{j}\right)\nonumber \\
\label{eq:formFactor}
\end{eqnarray}
is used to describe the screening effect in solids. For more details,
see Ref.~\citep{Liang:2020ryg}.

\subsection{\label{subsec:Renormalization}Electron-phonon interaction renormalization
at the RPA level}

\begin{figure}
\begin{centering}
\includegraphics[scale=0.5]{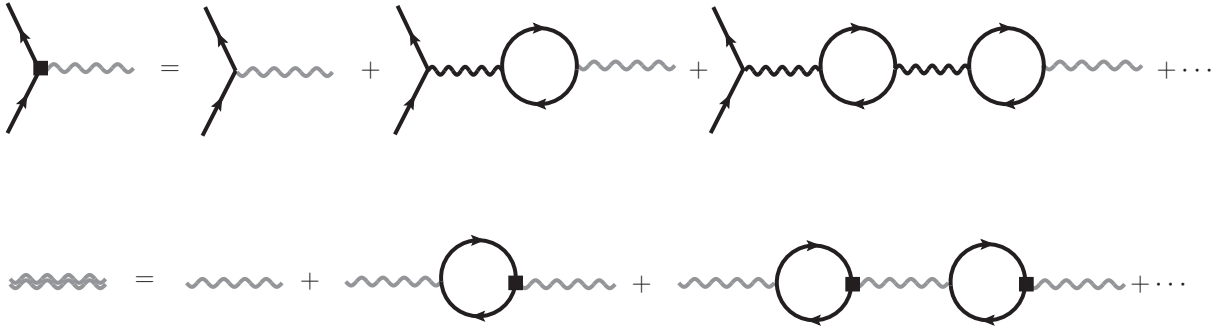}
\par\end{centering}
\caption{\label{fig:renormalization}\textbf{\textit{Top}}: The renormalized
electron-phonon vertex~(black square box) presented at the RPA level,
where the one-particle-irreducible blob is represented by an electron-hole
pair loop. Note that the gray lines represent the phonon propagators,
while the black wiggly lines represent the Coulomb interaction. \textbf{\textit{Bottom}}:
The dressed phonon line~(double wiggly line) presented at the RPA
level. See text for details.}
\end{figure}
Here we discuss how to renormalize both the bare phonon propagator
and effective electron-phonon interaction at the RPA level. In RPA,
the polarization bubble is approximated as a simple electron-hole
pair bubble, and hence the dressed electron-phonon vertex shown in
the top row of Fig.~\ref{fig:renormalization} can be expressed~(when
the reciprocal lattice vectors are suppressed) as the sum
\begin{eqnarray}
 &  & \frac{i\mathbf{\boldsymbol{\epsilon}}_{\mathbf{k},\alpha}\cdot\mathbf{k}}{\sqrt{N\,m_{N}}}\,\braket{i|e^{i\mathbf{k}\cdot\hat{\mathbf{x}}}|j}\,\left\{ 1+iv\left(\mathbf{k}\right)\left(-i\right)\Pi\left(\mathbf{k},\,\omega\right)+\left[iv\left(\mathbf{k}\right)\left(-i\right)\Pi\left(\mathbf{k},\,\omega\right)\right]^{2}+\cdots\right\} \left(-NZ_{\mathrm{ion}}\right)\frac{v\left(\mathbf{k}\right)}{V}\nonumber \\
 & = & \frac{i\mathbf{\boldsymbol{\epsilon}}_{\mathbf{k},\alpha}\cdot\mathbf{k}}{\sqrt{N\,m_{N}}}\,\braket{i|e^{i\mathbf{k}\cdot\hat{\mathbf{x}}}|j}\,\left[\frac{1}{1-v\left(\mathbf{k}\right)\Pi\left(\mathbf{k},\,\omega\right)}\right]\left(-NZ_{\mathrm{ion}}\right)\frac{v\left(\mathbf{k}\right)}{V}\nonumber \\
 & = & \frac{i\mathbf{\boldsymbol{\epsilon}}_{\mathbf{k},\alpha}\cdot\mathbf{k}}{\sqrt{N\,m_{N}}}\,\braket{i|e^{i\mathbf{k}\cdot\hat{\mathbf{x}}}|j}\,\frac{\left(-NZ_{\mathrm{ion}}\right)}{V}\,\frac{v\left(\mathbf{k}\right)}{\epsilon\left(\mathbf{k},\,\omega\right)},
\end{eqnarray}
where $\left(-i\right)\Pi$ corresponds to the electron-hole loop.
It is evident that the effect of the normalization is suppressing
the bare vertex with the dielectric function $\epsilon\left(\mathbf{k},\,\omega\right)$.
The electron-phonon interaction also bring about a correction to the
position of the phonon poles (in the RPA), as shown in the bottom
row of Fig.~\ref{fig:renormalization}. Since the typical band gaps
are much larger than the phonon eigenenergies, these small corrections
are irrelevant for our purpose in this study. One can use the $\mathtt{DarkELF}$
package to take into account the screening effect in various materials~\citep{Knapen:2021bwg}. 

\subsection{\label{subsec:Asymptotic}Asymptotic behavior of the multi-phonon
distribution}

Here we investigate the asymptotic behavior of the combined distribution
of $n$ independent Poisson distributions encountered in Sec.~\ref{sec:Multiphonon},
which justifies the validity of the impulse approximation in the large
$q$ regime. To make the discussion concise, the following parameters
are introduced: $\epsilon_{i}\sim\frac{\omega_{i}}{\overline{\omega}}$,
$\mu_{i}\sim\left(\frac{1}{3N}\right)\frac{1}{\epsilon_{i}}$, $\lambda\sim\frac{q^{2}}{2m_{N}\overline{\omega}}$,
$x\sim\frac{\omega}{\overline{\omega}\lambda}$, $\mu\sim\sum_{i=1}^{3N}\mu_{i}\epsilon_{i}=1$,
$\sigma^{2}\sim\sum_{i=1}^{3N}\mu_{i}\epsilon_{i}^{2}=1$, and then
consider the distribution
\begin{eqnarray}
P\left(n_{1},\,n_{2},\cdots,\,n_{3N}\right) & = & p\left(n_{1},\lambda\mu_{1}\right)p\left(n_{2},\lambda\mu_{2}\right)\cdots p\left(n_{3N},\lambda\mu_{3N}\right)\nonumber \\
 & = & \frac{\left(\lambda\mu_{1}\right)^{n_{1}}}{n_{1}!}e^{-\lambda\mu_{1}}\cdots\frac{\left(\lambda\mu_{3N}\right)^{n_{3N}}}{n_{3N}!}e^{-\lambda\mu_{3N}},
\end{eqnarray}
where $p\left(n_{i},\lambda\mu_{i}\right)$ is the Poisson distribution
of variable $n_{i}$ with mean $\lambda\mu_{i}$. We first try to
obtain the distribution of a random variable in the form of $z=\sum_{i=1}^{3N}\left(\epsilon_{i}/\lambda\right)n_{i}$,
and $\lambda$ being a parameter that characterizes the scale of the
problem. To achieve this goal, we calculate the relevant characteristic
function
\begin{eqnarray}
\varphi_{Z}\left(t\right) & = & \sum_{\left\{ n_{i}\right\} }P\left(n_{1},\,n_{2},\cdots,\,n_{3N}\right)e^{i\,zt}\nonumber \\
 & = & \prod_{i=1}^{3N}\left(\sum_{n_{i}=0}^{+\infty}\frac{\left(\lambda\mu_{i}\right)^{n_{i}}}{n_{i}!}e^{-\lambda\mu_{i}}e^{i\left(\epsilon_{i}/\lambda\right)n_{i}t}\right)\nonumber \\
 & = & \exp\left[\sum_{i=1}^{3N}\lambda\mu_{i}\left(e^{i\left(\epsilon_{i}/\lambda\right)t}-1\right)\right],
\end{eqnarray}
with which the distribution of the variable $x$ is explicitly expressed
and expanded in $\lambda\,\left(\lambda\gg1\right)$ as follows, 
\begin{eqnarray}
\Phi\left(x\right) & = & \frac{1}{2\pi}\int_{-\infty}^{+\infty}\varphi_{Z}\left(t\right)e^{-i\,xt}\mathrm{d}t\nonumber \\
 & = & \int_{-\infty}^{+\infty}\frac{\mathrm{d}t}{2\pi}\exp\left[\sum_{i=1}^{3N}\lambda\mu_{i}\left(e^{i\left(\epsilon_{i}/\lambda\right)t}-1\right)-i\,xt\right]\nonumber \\
 & = & \int_{-\infty}^{+\infty}\frac{\mathrm{d}t}{2\pi}\exp\left[-i\,(x-1)t-\left(\frac{1}{\lambda}\sum_{i=1}^{3N}\mu_{i}\epsilon_{i}^{2}\right)\frac{t^{2}}{2}\right]\cdot\left[1-i\left(\frac{1}{\lambda^{2}}\sum_{i=1}^{3N}\mu_{i}\epsilon_{i}^{3}\right)\frac{t^{3}}{3!}+o\left(\frac{1}{\lambda^{2}}\right)\right]\nonumber \\
 & = & \int_{-\infty}^{+\infty}\frac{\mathrm{d}z}{\sqrt{2\pi}}\exp\left\{ -\frac{1}{2}\left[z+i\sqrt{\lambda}\,\left(x-1\right)\right]^{2}\right\} \cdot\left[1-\frac{i}{\sqrt{\lambda}}\left(\sum_{i=1}^{3N}\mu_{i}\epsilon_{i}^{3}\right)\frac{z^{3}}{3!}+o\left(\frac{1}{\sqrt{\lambda}}\right)\right]\nonumber \\
 &  & \times\frac{1}{\sqrt{2\pi}/\sqrt{\lambda}}\exp\left[\frac{-\left(x-1\right)^{2}}{2/\lambda}\right]\nonumber \\
 & = & \frac{1}{\sqrt{2\pi}/\sqrt{\lambda}}\exp\left[\frac{-\left(x-1\right)^{2}}{2/\lambda}\right]\times\left\{ 1+\frac{1}{6}\left(\sum_{i=1}^{3N}\mu_{i}\epsilon_{i}^{3}\right)\left(x-1\right)\left[\lambda\left(x-1\right)^{2}-3\right]+o\left(\frac{1}{\sqrt{\lambda}}\right)\right\} ,\nonumber \\
\end{eqnarray}
where $\sum_{i=1}^{3N}\mu_{i}\epsilon_{i}^{3}=\sum_{i=1}^{3N}\left(\frac{1}{3N}\right)\left(\omega_{i}^{2}/\overline{\omega}^{2}\right)$
in the last line. The integral in the last step can be evaluated by
integrating over its saddle point $z_{0}=-i\sqrt{\lambda}\left(x-1\right)$
along the path $\left(z_{0}-\infty,\,z_{0}+\infty\right)$, on which
$z^{3}$ term is suppressed by $1/\sqrt{\lambda}$, as long as $\left|x-1\right|$
is not too much larger than the width $1/\sqrt{\lambda}$. 

\subsection{\label{subsec:Iterative_Multiphonon}Iterative calculation of multi-phonon
process}

Here we discuss how to calculate the multi-phonon spectrum at $T=0\,\mathrm{K}$
following a recursive procedure. First, we explicitly derive the scattering
factor introduced in Eq.~(\ref{eq:Compound}) as the following,
\begin{eqnarray}
S\left(q,\omega\right) & = & \sum_{\left\{ n_{i}\right\} }\frac{e^{-\frac{E_{R}\left(q\right)}{3N\omega_{1}}}}{n_{1}!}\left(\frac{E_{R}\left(q\right)}{3N\omega_{1}}\right)^{n_{1}}\cdots\frac{e^{-\frac{E_{R}\left(q\right)}{3N\omega_{3N}}}}{n_{3N}!}\left(\frac{E_{R}\left(q\right)}{3N\omega_{3N}}\right)^{n_{3N}}\delta\left(\sum_{i=1}^{3N}n_{i}\omega_{i}-\omega\right)\nonumber \\
 & = & e^{-E_{R}\left(q\right)\sum_{i=1}^{3N}\frac{1}{3N}\frac{1}{\omega_{i}}}\left[\int_{-\infty}^{+\infty}\frac{e^{i\left(\sum_{i=1}^{3N}n_{i}\omega_{i}-\omega\right)t}}{2\pi}\mathrm{d}t\right]\times\left[\sum_{\left\{ n_{i}\right\} }\frac{1}{n_{1}!}\left(\frac{E_{R}\left(q\right)}{3N\omega_{1}}\right)^{n_{1}}\cdots\frac{1}{n_{3N}!}\left(\frac{E_{R}\left(q\right)}{3N\omega_{3N}}\right)^{n_{3N}}\right]\nonumber \\
 & = & e^{-E_{R}\left(q\right)\sum_{i=1}^{3N}\frac{1}{3N}\frac{1}{\omega_{i}}}\int_{-\infty}^{+\infty}\mathrm{d}t\,\frac{e^{-i\omega t}}{2\pi}\times\left[\sum_{n=0}^{+\infty}\frac{E_{R}\left(q\right)^{n}}{n!}\left(\sum_{i=1}^{3N}\frac{1}{3N}\frac{e^{i\omega_{i}t}}{\omega_{i}}\right)^{n}\right]\nonumber \\
 & = & e^{-E_{R}\left(q\right)\sum_{i=1}^{3N}\frac{1}{3N}\frac{1}{\omega_{i}}}\int_{-\infty}^{+\infty}\mathrm{d}t\,\frac{e^{-i\omega t}}{2\pi}\times\left(\sum_{n=0}^{+\infty}\frac{E_{R}\left(q\right)^{n}}{n!}f\left(t\right)^{n}\right)\nonumber \\
 & = & e^{-E_{R}\left(q\right)\sum_{i=1}^{3N}\frac{1}{3N}\frac{1}{\omega_{i}}}\times\left(\sum_{n=0}^{+\infty}\frac{E_{R}\left(q\right)^{n}}{n!}\int_{-\infty}^{+\infty}\mathrm{d}t\,f\left(t\right)^{n}\frac{e^{-i\omega t}}{2\pi}\right)\nonumber \\
 & = & e^{-E_{R}\left(q\right)\sum_{i=1}^{3N}\frac{1}{3N}\frac{1}{\omega_{i}}}\times\left(\sum_{n=0}^{+\infty}\frac{E_{R}\left(q\right)^{n}}{n!}T_{n}\left(\omega\right)\right).
\end{eqnarray}
\begin{figure}
\begin{centering}
\includegraphics[scale=0.9]{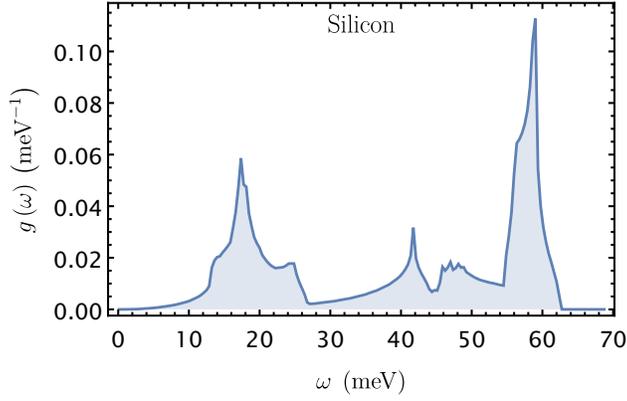}
\par\end{centering}
\caption{\label{fig:DOS}The normalized phonon density of states for bulk silicon.
See text for details.}
\end{figure}
Then it is easy to verify the following recursive relation~(see Ref.~\citep{Schober2014}
for a general discussion on the case of a finite temperature)
\begin{eqnarray}
\int_{-\infty}^{+\infty}T_{1}\left(\omega-\omega'\right)T_{p-1}\left(\omega'\right)\mathrm{d}\omega' & = & \int_{-\infty}^{+\infty}\mathrm{d}\omega'\int_{-\infty}^{+\infty}\mathrm{d}t\,f\left(t\right)\frac{e^{-i\left(\omega-\omega'\right)t}}{2\pi}\int_{-\infty}^{+\infty}\mathrm{d}t'\,f\left(t'\right)^{p-1}\frac{e^{-i\omega't'}}{2\pi}\nonumber \\
 & = & \frac{1}{2\pi}\int_{-\infty}^{+\infty}f\left(t\right)^{p}e^{-i\omega t}\mathrm{d}t\nonumber \\
 & = & T_{p}\left(\omega\right).\label{eq:recursion}
\end{eqnarray}
It is straightforward to see
\begin{eqnarray}
T_{1}\left(\omega\right) & = & \int_{-\infty}^{+\infty}\mathrm{d}t\,f\left(t\right)\frac{e^{-i\omega t}}{2\pi}\nonumber \\
 & = & \int_{-\infty}^{+\infty}\mathrm{d}t\,\frac{1}{3N}\sum_{i=1}^{3N}\frac{1}{\omega_{i}}\frac{e^{i\left(\omega_{i}-\omega\right)t}}{2\pi}\nonumber \\
 & = & \frac{1}{3N}\sum_{i=1}^{3N}\frac{1}{\omega}\delta\left(\omega_{i}-\omega\right),
\end{eqnarray}
which means $T_{1}\left(\omega\right)=0,$ if $\omega<0$. This feature
can be easily generalized to the case of an arbitrary $p$ such that
$T_{p}\left(\omega\right)=0$, $\left(\omega<0\right)$. In practice,
we utilize Eq.~(\ref{eq:recursion}) to obtain $\left\{ T_{n}\left(\omega\right)\right\} $
and to further calculate the spectrum of the multi-phonon process
with Eq.~(\ref{eq:Compound}). This recursive method requires only
the phonon DoS for a solid target (\textit{e.g.}, for monatomic simple
lattice) as follows,
\begin{eqnarray}
g\left(\omega\right) & = & \frac{1}{3N}\sum_{i=1}^{3N}\delta\left(\omega_{i}-\omega\right),
\end{eqnarray}
which is normalized such that $\int_{0}^{+\infty}g\left(\omega\right)\mathrm{d}\omega=1$.
For illustration, in Fig.~\ref{fig:DOS} we present the normalized
DoS for the bulk silicon, which is calculated using $\mathtt{PHONOPY}$
code~\citep{TOGO20151}, while the force constants are computed using
$\mathtt{VASP}$ package~\citep{PhysRevB.54.11169} based on the
density functional theory~\citep{PhysRev.136.B864,PhysRev.140.A1133}
with Perdew, Burke, and Ernzerhof form~\citep{PhysRevLett.77.3865}
of the generalized gradient approximation on the exchange-correlation
functional. 

\vspace{1cm}

\bibliographystyle{JHEP1}
\addcontentsline{toc}{section}{\refname}\bibliography{D:/Lyx/multiphonon/lyx_1/prd/prd2.0/SolarElectronScreening3}

\end{document}